\documentclass[fleqn,usenatbib]{mnras}
\usepackage{amsmath}	
\usepackage{mathptmx}
\usepackage{txfonts}
\usepackage[T1]{fontenc}
\usepackage{ae,aecompl}
\usepackage{amssymb}	
\usepackage{graphicx}	

\def\aj{AJ}                   
\def\apj{ApJ}                
\def\apjl{ApJ}                
\def\apjs{ApJS}               
\def\ao{Appl.~Opt.}           
\def\aap{A\&A}               
\def\aaps{A\&AS}              

\def\jcap{J. Cosmology Astropart. Phys.}
\def\mnras{MNRAS}             
\def\rmxaa{Rev. Mexicana Astron. Astrofis.}%
\def\nat{Nature}              

\def\physrep{Phys.~Rep.}   

\title[Integrated HI 21 cm Emission From Filaments]{Filament Hunting: Integrated HI 21 cm Emission From Filaments Inferred by Galaxy Surveys}

\author[Kooistra et al.]{Robin Kooistra,$^{1}$
Marta B. Silva$^{1}$
 and Saleem Zaroubi$^{1,2}$
\\
$^1$Kapteyn Astronomical Institute, University of Groningen, Landleven 12, 9747 AD Groningen, the Netherlands\\
$^2$Department of Natural Sciences, Open University of Israel, 1 University Road, PO Box 808, Ra'anana 4353701, Israel}
\begin{document}
\maketitle

\begin{abstract}
Large scale filaments, with lengths that can reach tens of Mpc, are the most prominent features in the cosmic web. These filaments have only been observed indirectly through the positions of galaxies in large galaxy surveys or through absorption features in the spectra of high-redshift sources. In this study, we propose to go one step further and directly detect intergalactic medium filaments through their emission in the HI 21 cm line. We make use of high-resolution cosmological simulations to estimate the intensity of this emission in low-redshift filaments and use it to make predictions for the direct detectability of specific filaments previously inferred from galaxy surveys, in particular the Sloan Digital Sky Survey. Given the expected signal of these filaments, our study shows that HI emission from large filaments can be observed by current and next-generation radio telescopes. We estimate that gas in filaments of length $l \gtrsim$ 15 $h^{-1}$Mpc with relatively small inclinations to the line of sight ($\lesssim 10^\circ$) can be observed in $\sim100$ h with telescopes such as Giant Metrewave Telescope or Expanded Very Large Array, potentially providing large improvements over our knowledge of the astrophysical properties of these filaments. Due to their large field of view and sufficiently long integration times, upcoming HI surveys with the Apertif and Australian Square Kilometre Array Pathfinder instruments will be able to detect the brightest independently of their orientation and curvature. Furthermore, our estimates indicate that a more powerful future radio telescope like Square Kilometre Array can be used detect even the faintest of these filaments with integration times of $\sim10-40$ h.

\end{abstract}

\begin{keywords}
intergalactic medium -- cosmology: theory -- diffuse radiation -- large-scale structure of Universe 
\end{keywords}
 
\section{Introduction}
Observations of the local Universe have revealed that most galaxies reside in a complex network of filamentary structures 
known as the cosmic web \citep{art:web}. Within the ${\Lambda}$ cold dark matter framework, these structures are an expected result from nonlinear gravitational evolution. According to this picture, the dark matter haloes within which galaxies reside are connected to each other through a patchwork of filaments and sheets that constitute the structure of the intergalactic medium \citep[IGM; e.g.,][]{art:cen99}. Whereas galaxies reside in high density regions, the diffuse gaseous component, if observed, gives an alternative view into filaments and their baryonic component. Furthermore, it
potentially gives a more faithful tracer to the spatial distribution of the IGM than galaxies by themselves, which in turn allows for the study of the connection between  galaxies and the filaments of gas.\\

Outside galaxies most baryons are expected to be in the relatively dense circumgalactic 
medium, which is mostly ionized and warm to hot. The remaining gas is in a filamentary structure with gas heated and ionized 
due to gravitational collapse and the UV and X-ray backgrounds. However, shielding from ionizing radiation  due to recombinations in the denser 
pockets of gas still allows some of this gas to have a higher neutral fraction and emit potentially observable HI 21 cm radiation. Generally, the neutral fraction in 
these regions is well below a percent of the total gas content \citep{art:popping,art:takeuchi}. Nonetheless, its existence 
has been proved observationally through Lyman alpha forest absorption in the spectra of high redshift quasars \citep{art:GP}. 
However, meaningful Lyman alpha absorption requires neutral gas above a given density threshold. Moreover, the technique is limited by the 
lines of sight to the available quasars and the matter content inferred from it depends on several assumptions about the gas 
conditions that are deduced from comparisons with simulations \citep[e.g.,][]{art:borde}.\\

Thus far, large scale filaments have mostly been detected through tracing the spatial distribution of galaxies in large galaxy 
surveys, such as the Sloan Digital Sky Survey \citep[SDSS, ][]{art:sdss1,art:sdss8}, the 2-degree Field (2dF) Galaxy Redshift 
Survey \citep{art:2df} and the Two Micron All-Sky Survey \citep[2MASS, ][]{art:2mass}. In such surveys the larger structures of the cosmic web 
can be easily identified. Indeed, some effort has been made to catalogue filaments traced by SDSS galaxies and many filaments with lengths 
ranging from a few to tens of $h^{-1}$Mpc and diameters of $\sim$0.5-2 $h^{-1}$Mpc have been found 
\citep[e.g.,][]{art:tempel,art:sousbie08,art:jasche10,art:smith12}.

In recent years, some effort have been devoted to tracing the large scale structure of the cosmic web through the gas distribution.
In a ground braking work, \citet{art:chang10} used HI 21 cm intensity maps obtained from the Green Bank Telescope (GBT), to constrain the large scale structure at a redshift of z 
$\sim0.8$ by cross-correlating with data from galaxy surveys. The same group followed this work up with a more significant detection
using the WiggleZ galaxy survey data together with GBT 21cm data \citep{art:masui13}. 
At higher redshifts, evidence for filamentary structures was found in the spectra of background sources
due to scattering of Lyman alpha photons by the neutral IGM along the line of sight \citep[e.g.,][]{art:abs1,art:abs2,art:abs3,art:finley14}. 
Recently, a cosmic web filament at z $\sim$ 2-3 illuminated by a bright quasar was detected due to its Lyman alpha fluorescent emission \citep{art:fildet}. 
At lower redshifts, however, the gas in the IGM is on average more ionized and the Lyman alpha line is observed in the UV, and 
so it is much more challenging to detect it in emission.\\

\citet{art:takeuchi} explored the possibility of using the HI 21 cm line to directly observe IGM filaments. By estimating the integrated HI 21 cm line intensity from simulated filaments they found that filaments with a length of $\sim100\, {\rm Mpc}$ can be detected in a 100 hours of integration by reasonably sensitive telescopes, such as the Giant Metrewave Radio Telescope (GMRT) or the Five hundred meter Aperture Spherical Telescope (FAST). They also found that the signal from filaments aligned along the line of sight can be more easily detected. Obviously, detection of such a signal would allow for the study of both the baryonic content of the filament and the ionization state of the gas. This would in turn help constrain the UV background, since this radiation is the main agent responsible for the thermal and ionization state of the gas.\\
Recently, \citet{art:horii17} performed a similar exercise, but then for the warm hot intergalactic medium (WHIM). Their simulations include strong feedback, resulting in very high temperature filaments. This in turn leads to a large ionization fraction and an HI 21 cm signal that is more difficult to observe.\\

In this work we go one step further by directly linking the properties of cosmic web filaments in our simulations to those that have previously been identified from observational galaxy catalogues. Specifically, we use the filament catalogue by \citet{art:tempel}, obtained from SDSS data, to find real filaments that are aligned along the line of sight and 
extract similar filaments from the simulation.
 With this identification we then employ the simulations to calculate the 
neutral gas fraction in these specific filaments and predict their observability 
by current and future radio telescopes. 
To do this, we use a high resolution cosmological n-body simulation and apply a semi-analytical prescription to the density field to estimate the 
temperature and ionization state of the IGM, from which we determine the HI 21 cm signal from these filaments. 

The content of the paper is organized as follows. We begin by describing the model that is used to determine the thermal and 
ionization states of the IGM in Section \ref{secmodel}. The method to calculate the 21cm brightness temperature signal is 
shown in Section \ref{sec:sig}. The simulation details are presented in Section \ref{sec:sim}. We then 
describe how we selected the observed filaments and how we estimated their 
integrated HI 21 cm signal in Section \ref{sec:fils}. In the same section we introduce the instruments that are considered for possible observations 
and compare their sensitivity to the signals of the filaments. We explore the advantages of survey instruments with large field of view in more detail in Section \ref{sec:surveys} and finally the removal of contamination due to emission from galaxies in the HI 21 cm line 
is discussed in Section \ref{sec:contam}.\\
Throughout this work, we assume the \citet{art:planck15para} cosmological parameters ($\Omega_{\mathrm{m}}$ = 0.3089, 
$\Omega_{\mathrm{\Lambda}}$ = 0.6911, $\Omega_{\mathrm{b}}h^{2}$ = 0.02230, $H_{\rm{0}}$ = 67.74 $\mathrm{km\,s^{-1}Mpc^{-1}}$ and $Y_{\rm{P}}$ = 0.249).

\section{Model: Ionization and thermal state in the IGM}\label{secmodel}
The conditions in the IGM drive the observed HI signal. This depends on the complex interplay 
between the different ionization, recombination, heating and cooling processes in the gas. 
The thermal and ionization state of the cold IGM is mainly set by the strength of the UV/X-ray background and  
can be estimated by assuming both thermal and ionization equilibrium. The equilibrium assumptions are a good approximation 
for most of the gas in filaments, given that the relevant timescales for ionization and recombination are 
relatively short. However, they break down in very low density regions, where the gas cannot 
efficiently cool through recombination and collisional emission, and in the vicinity of galaxies or active galactic 
nuclei, where the ionizing radiation is much stronger and therefore the gas is highly ionized.
Here we describe the full details of the 
model used to determine the ionization and thermal states of the IGM.

The ionization state of the intergalactic medium is governed by the balance between ionization and 
recombination processes. Even at low redshift, the gas in filaments far from local sources is expected to be 
very metal poor and so the cooling and heating processes in this medium are dominated by reactions involving only 
hydrogen and helium. The fractions of the different states of hydrogen and helium can be found by solving the 
following set of balance equations \citep{art:fukkaw}.

\begin{figure*} 
\begin{centering}  
\includegraphics[angle=0,width=0.48\textwidth]{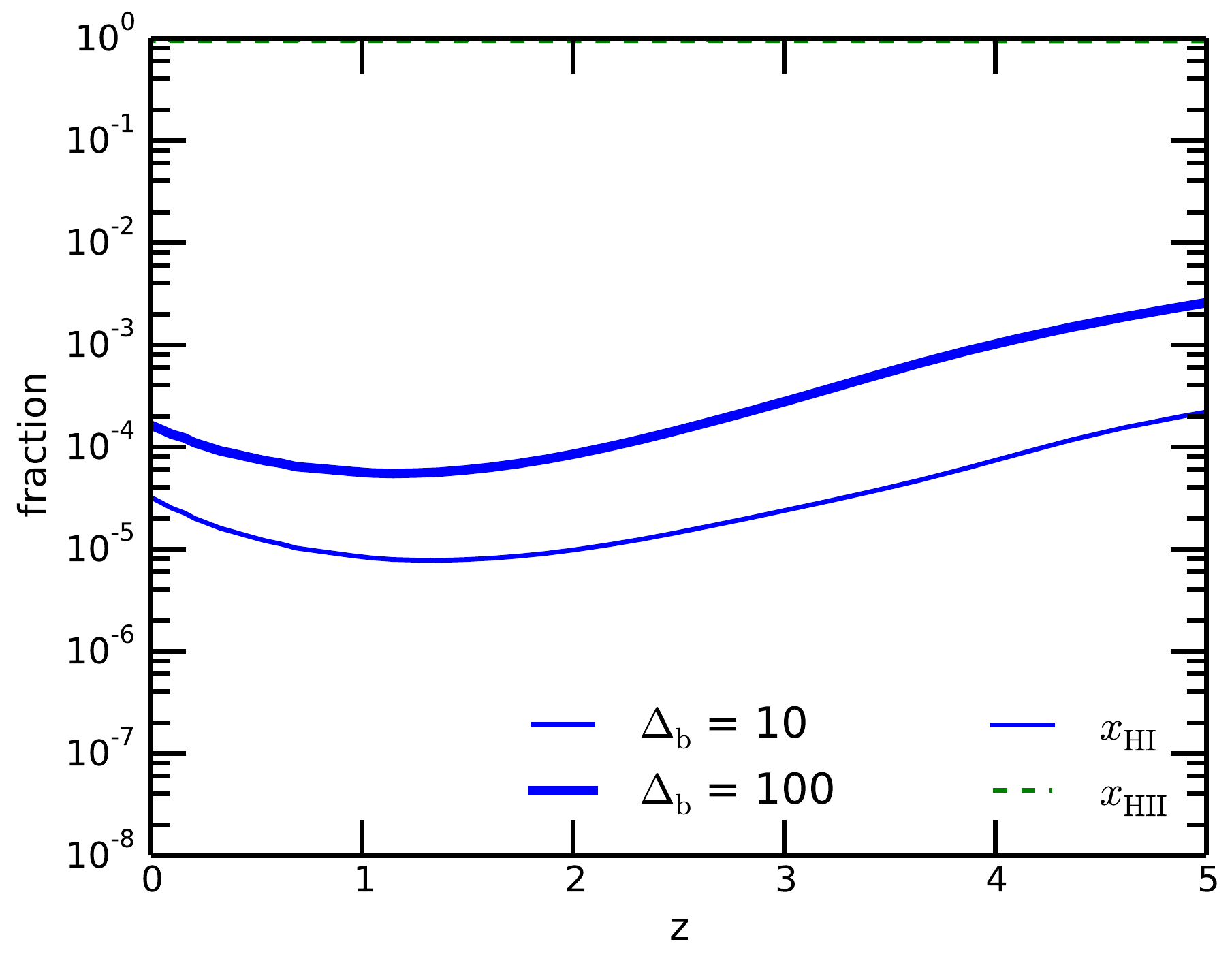}\quad
\includegraphics[angle=0,width=0.48\textwidth]{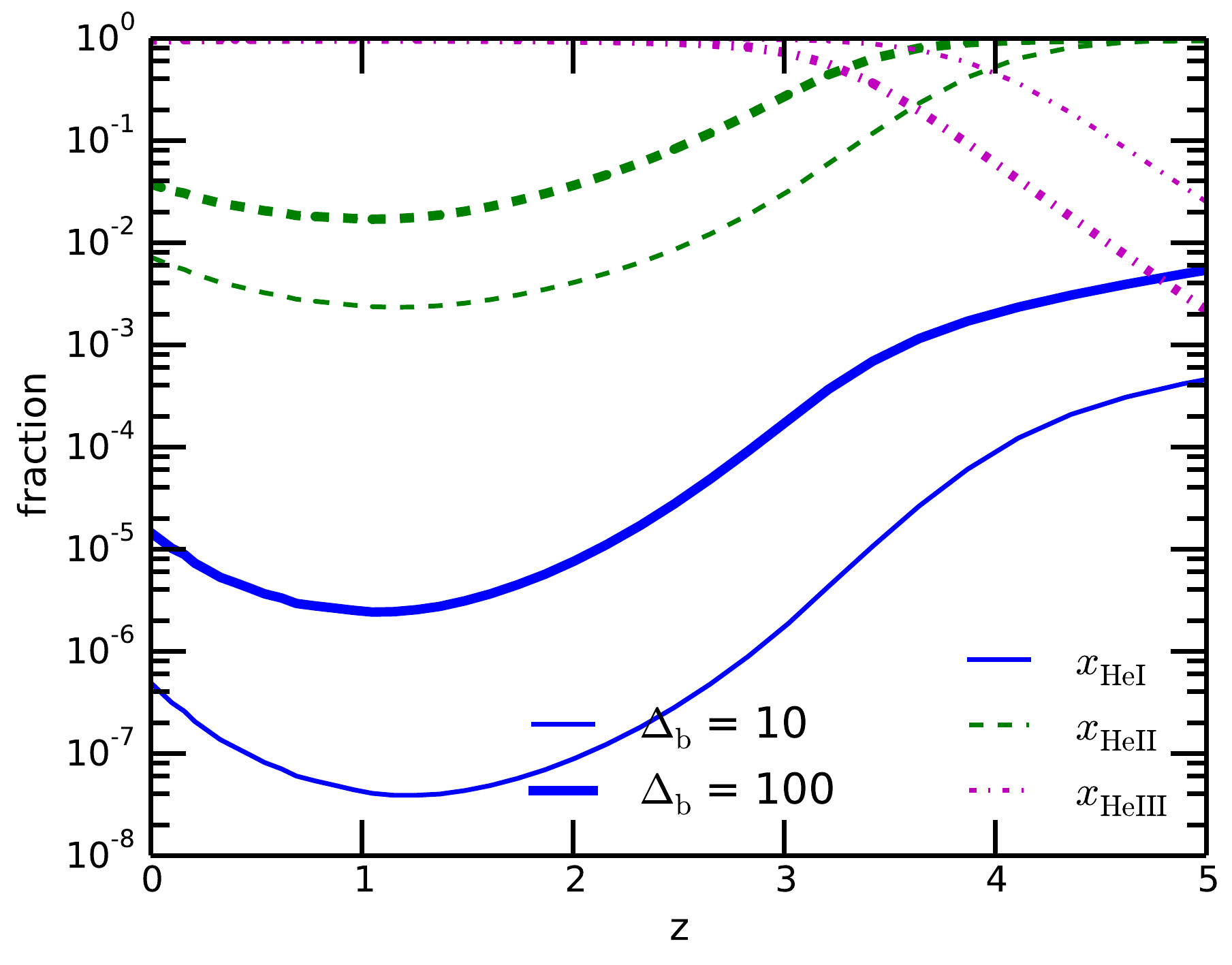}
\caption{Evolution of the ionization fractions for hydrogen (left) and helium (right). The width of the line denotes the density for which the calculation was performed.}
\label{fig:fractions}
\end{centering}
\end{figure*} 

\begin{align}
\frac{\mathrm{d}x_{\rm{HII}}}{\mathrm{d}t} & = \Gamma_{\rm{HI}}x_{\rm{HI}} + \beta_{\rm{HI}}n_{\mathrm{e}}x_{\rm{HI}} - \alpha_{\rm{HII}}n_{\mathrm{e}}x_{\rm{HII}}\label{rate:HII},\\
\frac{\mathrm{d}x_{\rm{HeII}}}{\mathrm{d}t} & = \Gamma_{\rm{HeI}}x_{\rm{HeI}} + \beta_{\rm{HeI}}n_{\mathrm{e}}x_{\rm{HeI}}\notag\\
   &\quad-\left(\alpha_{\rm{HeII}}+\xi_{\rm{HeII}}\right)n_{\mathrm{e}}x_{\rm{HeII}}\notag\\
   &\quad- \beta_{\rm{HeII}}n_{\mathrm{e}}x_{\rm{HeII}} - \Gamma_{\rm{HeII}}x_{\rm{HeII}}\notag\\
   &\quad+ \alpha_{\rm{HeIII}}n_{\mathrm{e}}x_{\rm{HeIII}},\label{rate:HeII}\\
\frac{\mathrm{d}x_{\rm{HeIII}}}{\mathrm{d}t} & = \Gamma_{\rm{HeII}}x_{\rm{HeII}} + \beta_{\rm{HeII}}n_{\mathrm{e}}x_{\rm{HeII}}\notag\\
   &\quad-\alpha_{\rm{HeIII}}n_{\mathrm{e}}x_{\rm{HeIII}}.\label{rate:HeIII}
\end{align}
Here $n_{\mathrm{e}}$ is the electron number density, whereas $x_{\rm{HI}}$, $x_{\rm{HII}}$, $x_{\rm{HeI}}$, $x_{\rm{HeII}}$ and $x_{\rm{HeIII}}$ 
denote the fractions of HI, HII, HeI, HeII and HeIII, respectively. $\Gamma_i$ is the photoionization 
rate, $\beta_i$ the collisional excitation rate and $\alpha_i$ the recombination rate of species 
$i$. $\xi_{\mathrm{HeII}}$ is the dielectronic recombination rate of HeII. For the photoionization 
rates we interpolate the tables of the ionizing background from \citet{art:hm12}. The recombination 
and collisional rates were determined using known temperature dependent parameterisations (see Appendix \ref{app:recion}).

From energy conservation of the IGM in an expanding universe, the gas temperature $T_{\mathrm{g}}$ follows
\begin{equation}
 \frac{dT_{\mathrm{g}}}{dt} = -2H(z)T_{\mathrm{g}} + \frac{2}{3}\frac{\left(\mathcal{H} - \Lambda\right)}{nk_{\mathrm{B}}}, \label{eq:tg}
\end{equation}
where $H$(z) denotes the Hubble parameter, $\mathcal{H}$ the heating rate, $\Lambda$ the cooling 
function and $n$ the baryon number density defined as $n\equiv n_H + 4n_{He}$. The first term on 
the right-hand side accounts for the adiabatic cooling due to the Hubble expansion of the Universe. 
For the heating function we adopt the values corresponding to the \citet{art:hm12} ionizing background. 
Our cooling function includes collisional ionization/excitation, (dielectronic) recombination, free-free 
emission and Compton scattering of the CMB photons. A detailed description of the adopted cooling rates 
can be found in Appendix \ref{app:cool}.

\begin{figure} 
\begin{centering}  
\includegraphics[angle=0,width=0.48\textwidth]{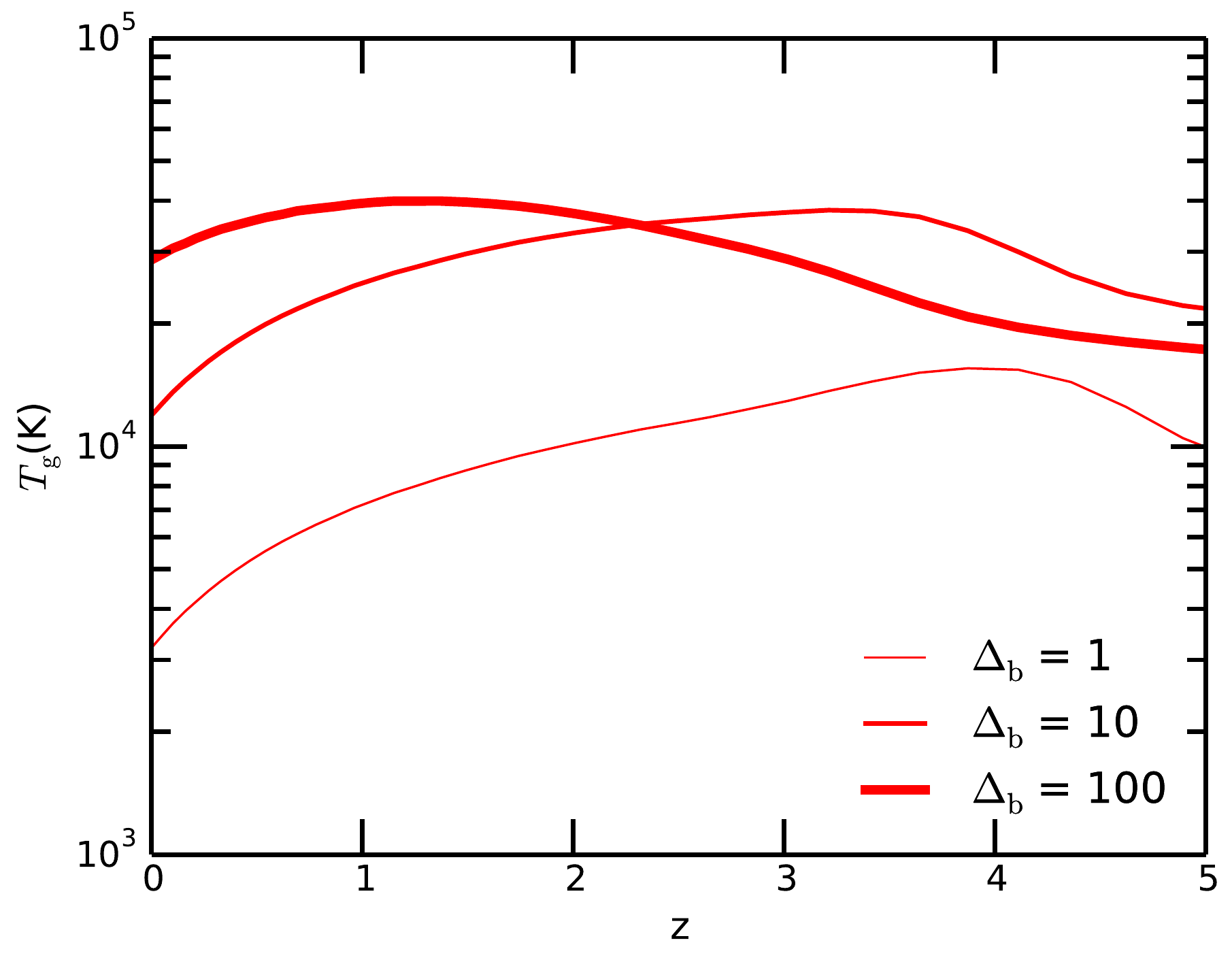}
\caption{Redshift evolution of the gas temperature. The width of the lines denotes the overdensity.}
\label{fig:gastemp}
\end{centering}
\end{figure}

As mentioned above, we solve these equations assuming ionization and thermal equilibrium, namely, that the LHS of the equations
is zero.
Due to the dependence of Equations \ref{rate:HII}, \ref{rate:HeII} 
and \ref{rate:HeIII} on the gas temperature, they need to be solved together with Equation \ref{eq:tg}, 
which we do iteratively. Our code solves for the ionization fractions and the gas temperature for a 
given hydrogen density and redshift. The redshift evolution of the ionization fractions of hydrogen 
and helium for different densities, relevant for filaments ($\Delta_\mathrm{b} \equiv \rho_{\rm b}/\left<\rho_{\rm b}\right>$), 
are shown in Figure \ref{fig:fractions} and the evolution of the gas temperature in Figure \ref{fig:gastemp}. 
The gas temperature will thus be $\sim 1-3\times 10^4$ K for most of the gas in a filament. At redshift z = 3.5 our 
temperature-density distribution matches the median equilibrium solution in the hydrodynamical 
simulations with radiative transfer by \citet{art:puchwein}. These simulations also use the evolving \citet{art:hm12} UV background, 
and were shown to reproduce the IGM temperature as predicted by Lyman alpha forest observations. The computed 
neutral hydrogen fraction and gas temperature are stable for small variations of the ionizing background. For overdensities of 100 and higher, the HI fraction would be high enough to make a significant contribution to the cosmological mass density of neutral hydrogen $\Omega_{\mathrm{HI}}$ \citep[e.g., Figure 12 in][]{art:crighton15}.

\section{HI 21 cm Brightness temperature signal}\label{sec:sig}
Instead of directly measuring the intensity, radio telescopes measure the contrast between the brightness of the observed object and that of the CMB. This signal is expressed as the differential brightness temperature and is given by
\begin{align}
 \delta T_{\mathrm{b}}^X (z) = &T_{\mathrm{b}}^X(z) - T_{\gamma}(z)\notag\\
  = &\frac{\left[T_{\mathrm{s}}^X(z) - T_{\gamma}(z)\right]\cdot\left(1-e^{-\tau_X(z)}\right)}{1+z},
\end{align}
where $T_\gamma$ is the CMB temperature and $T_{\rm s}^X$ and $\tau_X$ are the spin temperature and the optical depth of species X (i.e. HI), respectively. 
So if the spin temperature is lower than that of the CMB, the signal will appear in absorption and if it is higher, it will be in emission. 
In general, the optical depth is given by \citep{art:furlanetto06}
\begin{equation}
 \tau_X(z) = \frac{g_1}{g_0 + g_1}\frac{c^3A_{10}}{8\pi\nu_{10}^3}\frac{h_\mathrm{p}\nu_{10}}{k_{\mathrm{B}}T_{\mathrm{s}}}\frac{n_X(z)}{1+z}\frac{1}{dv_\parallel/dr_\parallel},
\end{equation}
where $h_{\mathrm{p}}$ is the Planck constant, $A_{10}$ is the transition probability, $k_\mathrm{B}$ the Boltzmann constant, $g_0$ and $g_1$ are the statistical 
weights of the ground and excited states, $\nu_{10}$ is the frequency at which the hyperfine transition occurs (for neutral 
hydrogen $A_{10} = 2.867\times10^{-15} \mathrm{s}^{-1}$, $g_1/g_0 = 3/1$ and $\nu_{10} = 1420.4\,\mathrm{MHz}$),  $n_X(z)$ is the physical number density of species X 
and $dv_\parallel/dr_\parallel$ is the comoving radial velocity gradient along the line of sight. When including peculiar 
velocities, the latter is given by $dv_\parallel/dr_\parallel = 1/(1+z)\left[H(z) + dv_r/dr\right]$, where $dv_r/dr$ 
is the comoving gradient of the line of sight component of the comoving velocity. In the optically thin limit, the 
differential brightness temperature becomes
\begin{align}
\delta T_{\mathrm{b}}^X (z) =& \frac{g_1}{g_0 + g_1}\frac{c^3h_{\mathrm{p}} A_{10}}{8\pi k_{\mathrm{B}}\nu_{10}^2}\frac{n_X(z)}{\left(1+z\right)H(z)}\notag\\
&\left(1-\frac{T_\gamma(z)}{T_s}\right)\left[1 + H(z)^{-1}dv_r/dr\right]^{-1}.\label{eq:dtb}
\end{align}
So the signal depends on the general properties of the line, the density of the medium and the cosmology, 
but it is then modified by the peculiar velocity term of order unity and by the spin temperature term. 
The latter becomes important when the spin temperature is close to the CMB temperature. In general, in 
studies of galaxies, the spin temperature is much higher than $T_{\gamma}$ and is therefore safe to ignore. 
However, for the lower density IGM that is not always the case and so it needs to be properly estimated.

\subsection{Spin temperature}
\begin{figure} 
\begin{centering}  
\hspace{-2 mm}
\includegraphics[angle=0,width=0.48\textwidth]{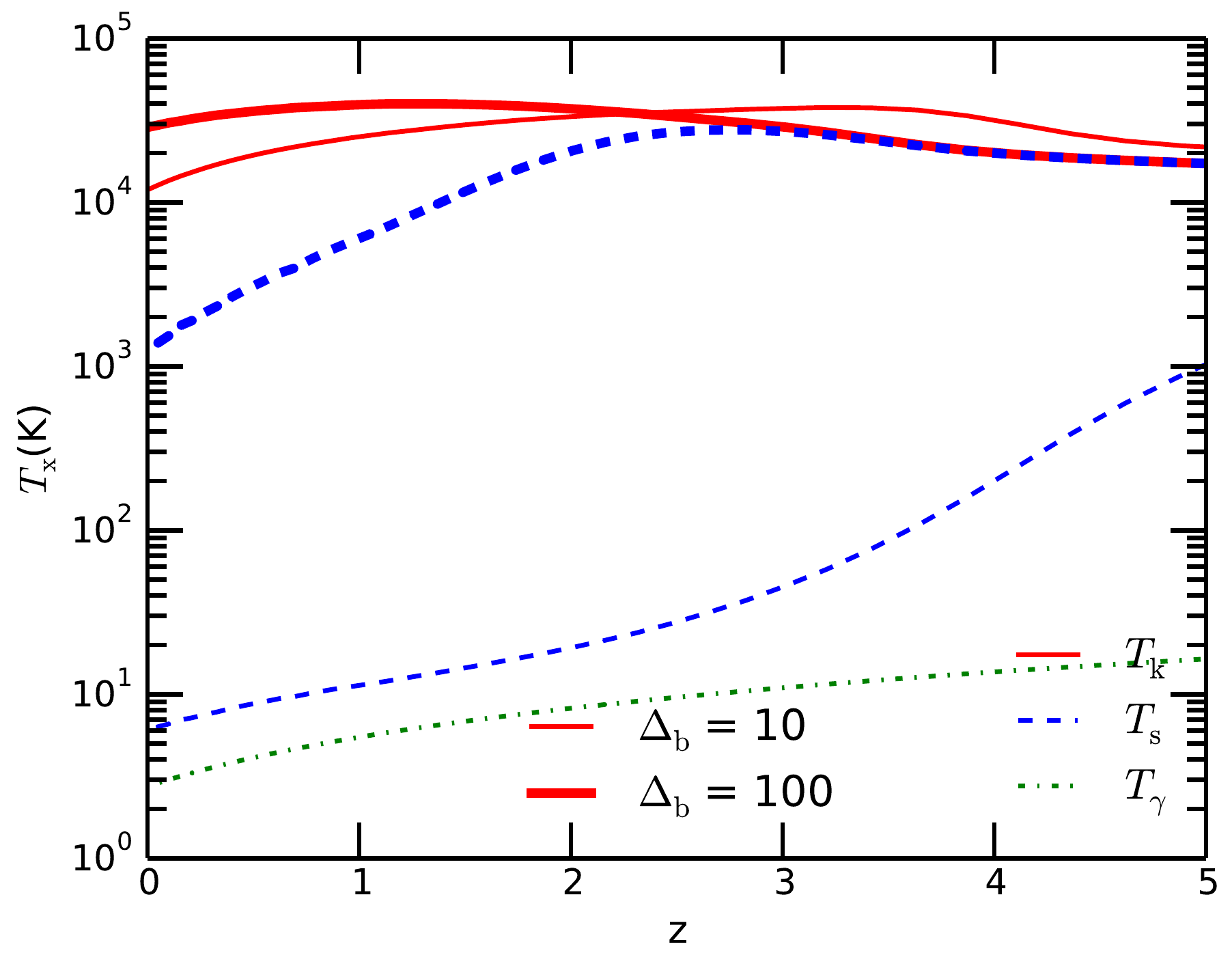}
\caption{Spin temperature evolution of hydrogen compared to the CMB temperature and kinetic temperature. The linewidth denotes the density.}
\label{fig:tspin}
\end{centering}
\end{figure}

The spin temperature $T_s$ determines the relative abundance of the exited state versus the ground state 
through the Boltzmann equation as
\begin{equation}
 \frac{n_1}{n_0} = \frac{g_1}{g_0}\exp\left(-\frac{h_{\mathrm{p}}\nu_{10}}{k_{\mathrm{B}} T_{\mathrm{s}}}\right),
\end{equation}
where $n_1$ and $n_0$ are the number of particles in the excited and ground state, respectively. The spin 
temperature is governed by absorption of CMB photons, collisions with hydrogen atoms, free electrons and 
protons and by scattering of UV photons (Wouthuysen-Field effect, \citet{art:WFw,art:field58}), which couple 
the spin temperature to the CMB photons and to the gas \citep{art:furlanetto06}. Therefore, the spin 
temperature can be written as \citep{art:field58}
\begin{equation}
 T_{\mathrm{s}}^{-1} = \frac{T_\gamma^{-1} + x_{\mathrm{c}} T_{\mathrm{k}}^{-1} + x_\alpha T_\alpha^{-1}}{1 + x_{\mathrm{c}} + x_\alpha},
\end{equation}
with $T_{\mathrm{k}}$ the kinetic temperature of the gas and $T_\alpha$ 
the colour temperature. The collisional coupling factor $x_{\mathrm{c}}$ and the Wouthuysen-Field 
coupling factor $x_\alpha$ are given by 
\begin{equation}
 x_{\mathrm{c}} = \frac{C_{10}}{A_{10}}\frac{T_*}{T_\gamma},\qquad x_\alpha = \frac{P_{10}}{A_{10}}\frac{T_*}{T_\gamma},
\end{equation}
where $A_{10}$ is the spontaneous decay rate from state 1 to state 0, which for neutral hydrogen has 
a value of $2.867\times10^{-15} \mathrm{s}^{-1}$, $C_{10}$ is the collisional de-excitation rate and 
$P_{10}$ is the de-excitation rate due to absorption of a Lyman alpha photon. The equivalent 
temperature $T_*$ is defined as $T_*\equiv h_{\mathrm{p}}\nu_{10}/k_{\mathrm{B}}$. When calculating 
the spin temperature, we assume that the kinetic and colour temperatures follow the gas temperature: 
$T_\alpha \sim T_{\mathrm{k}} \sim T_{\mathrm{g}}$.
The collisional de-excitation rate for neutral hydrogen can be expressed as a sum over the collisional 
processes with electrons, protons and other neutral hydrogen atoms. 
\begin{equation}
 C_{10}^{\mathrm{HI}} = \kappa_{10}^{\mathrm{HH}}(T_{\mathrm{k}})n_{\mathrm{H}} + \kappa_{10}^{\mathrm{eH}}(T_{\mathrm{k}})n_{\mathrm{e}} + \kappa_{10}^{\mathrm{pH}}(T_{\mathrm{k}})n_{\mathrm{p}}
\end{equation}
Here $\kappa_{10}^{\mathrm{HH}}$, $\kappa_{10}^{\mathrm{eH}}$ and $\kappa_{10}^{\mathrm{pH}}$ denote 
the collision rates for each process. Expressions for these can be found in \citet{art:zygelman}, 
\citet{art:sigfurl} and \citet{art:furlfurl,art:furlfurl2}\\
The de-excitation rate for the Wouthuysen-Field effect equals
\begin{equation}
 P_{10}^{\mathrm{HI}} = \frac{16\pi e^2 f_\alpha^{\mathrm{HI}}}{27m_{\mathrm{e}}c}J_{\mathrm{Ly\alpha,HI}},
\end{equation}
where $e$ is the electron charge, $f_\alpha$ the oscillator length of the Lyman alpha transition 
($f_\alpha = 0.4162$ for neutral hydrogen), $m_{\mathrm{e}}$ the electron mass and $c$ the speed 
of light. For the Lyman alpha photon angle-averaged specific intensity $J_{\mathrm{Ly\alpha,HI}}$ 
a model of the Lyman alpha emission that interacts locally with the IGM is required. We assume 
three sources of Lyman alpha photons: collisional excitations, recombinations and high energy 
background photons that redshift into the the Lyman alpha line. The full details of our calculation 
can be found in Appendix \ref{app:lya} and the resulting evolution of the spin temperature is 
shown in Figure \ref{fig:tspin}, where we also show the evolution of the gas temperature and the 
CMB temperature. For the high densities the spin temperature quickly couples to the kinetic 
temperature, but at low densities and low redshift the spin temperature approaches the CMB 
temperature, therefore suppressing the brightness temperature signal.

\subsection{Test: slab model}
\begin{figure} 
\begin{centering} 
\hspace{-2 mm}
\includegraphics[angle=0,width=0.48\textwidth]{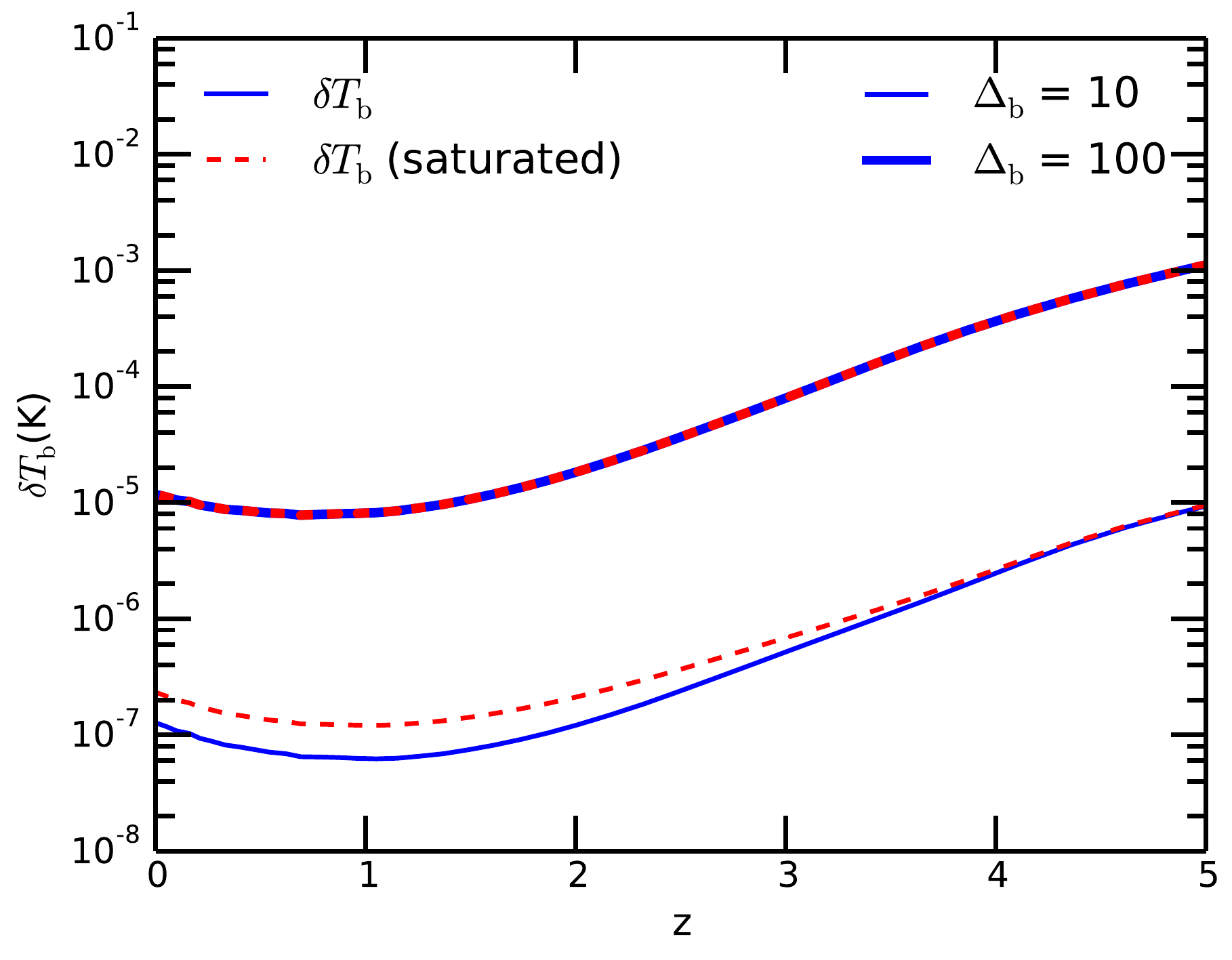}
\caption{Differential brightness temperature evolution for the slab model. The density is denoted by the width of the lines and the dashed lines show the signal in the saturated limit $\left(\left[1-T_{\gamma}/T_{\rm s}\right] \sim 1\right)$.}
\label{fig:dtbslab}
\end{centering}
\end{figure}

To test the effect of the spin temperature correction term on the resulting signal of a filament, 
we assume a simple constant density slab model. The signal from such a slab is calculated through 
a small adjustment of Equation \ref{eq:dtb}, giving \citep{art:takeuchi}

\begin{equation}
 \delta T_{\mathrm{b}}^X (z) = \frac{g_1}{g_0 + g_1}\frac{c^3h_{\mathrm{p}}A_{10}}{8\pi k_{\mathrm{B}}\nu_{10}^2}\frac{n_X(z)}{\left(1+z\right)^2}\left(1-\frac{T_\gamma(z)}{T_{\mathrm{s}}}\right)\frac{\Delta r}{\Delta v}
\end{equation}
with the width of the slab $\Delta r$ = 1 Mpc $h^{-1}$ and its proper line of sight velocity $\Delta v$ = 300 km $\rm{s^{-1}}$. 
The resulting differential brightness temperature evolution of HI is given in Figure \ref{fig:dtbslab}. The red 
dashed lines show the signal in the saturated limit $\left(\left[1-T_\gamma/T_\mathrm{s}\right] \sim 1\right)$ and the blue 
solid lines show the corrected signal. As can be seen, the spin temperature correction becomes significant with a 
factor of a few for the lower density filaments at low redshifts and therefore needs to be included in these calculations.
 
\begin{figure*} 
\begin{centering}  
\includegraphics[angle=0,width=0.49\textwidth]{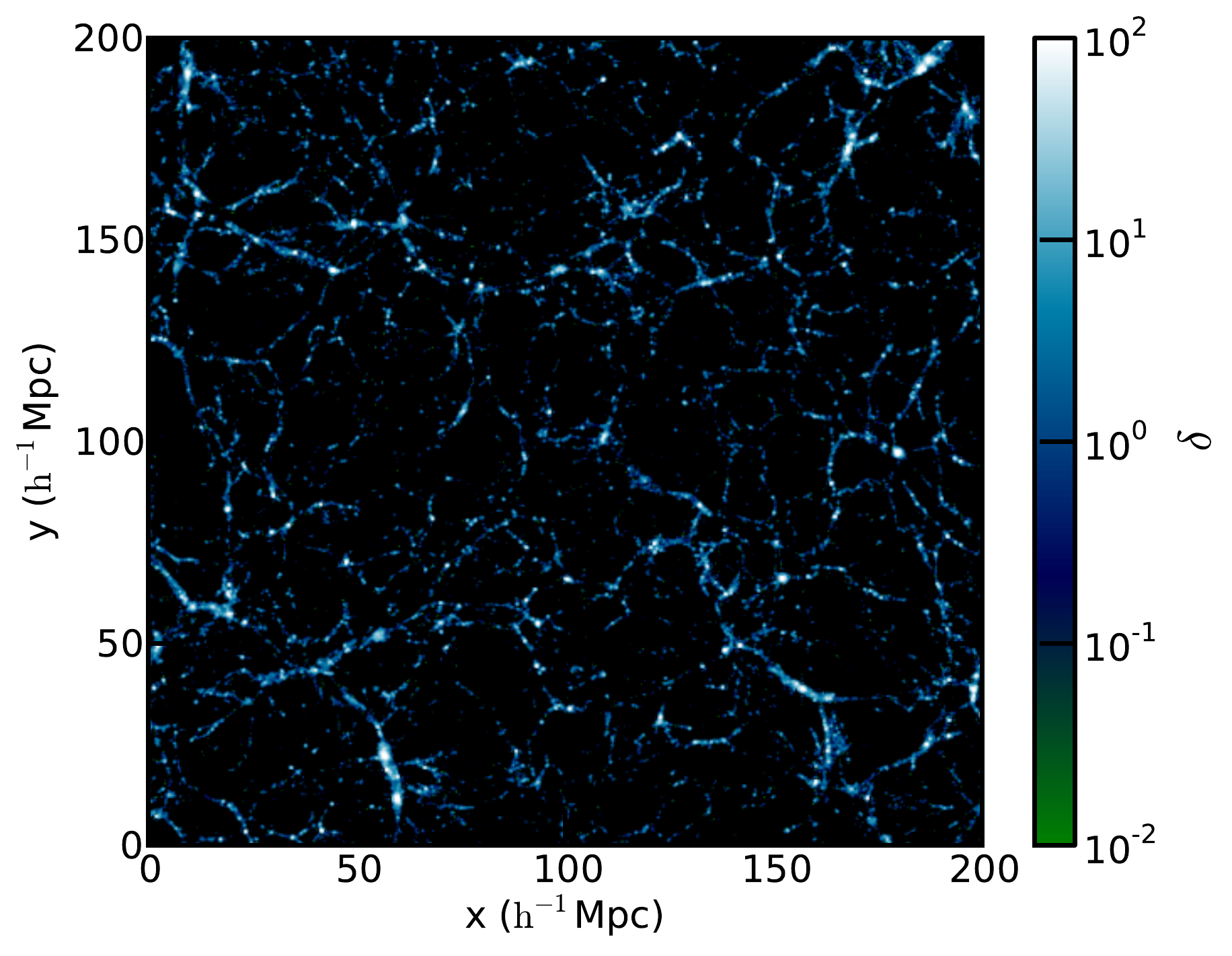}\quad
\hspace{-3 mm}
\includegraphics[angle=0,width=0.50\textwidth]{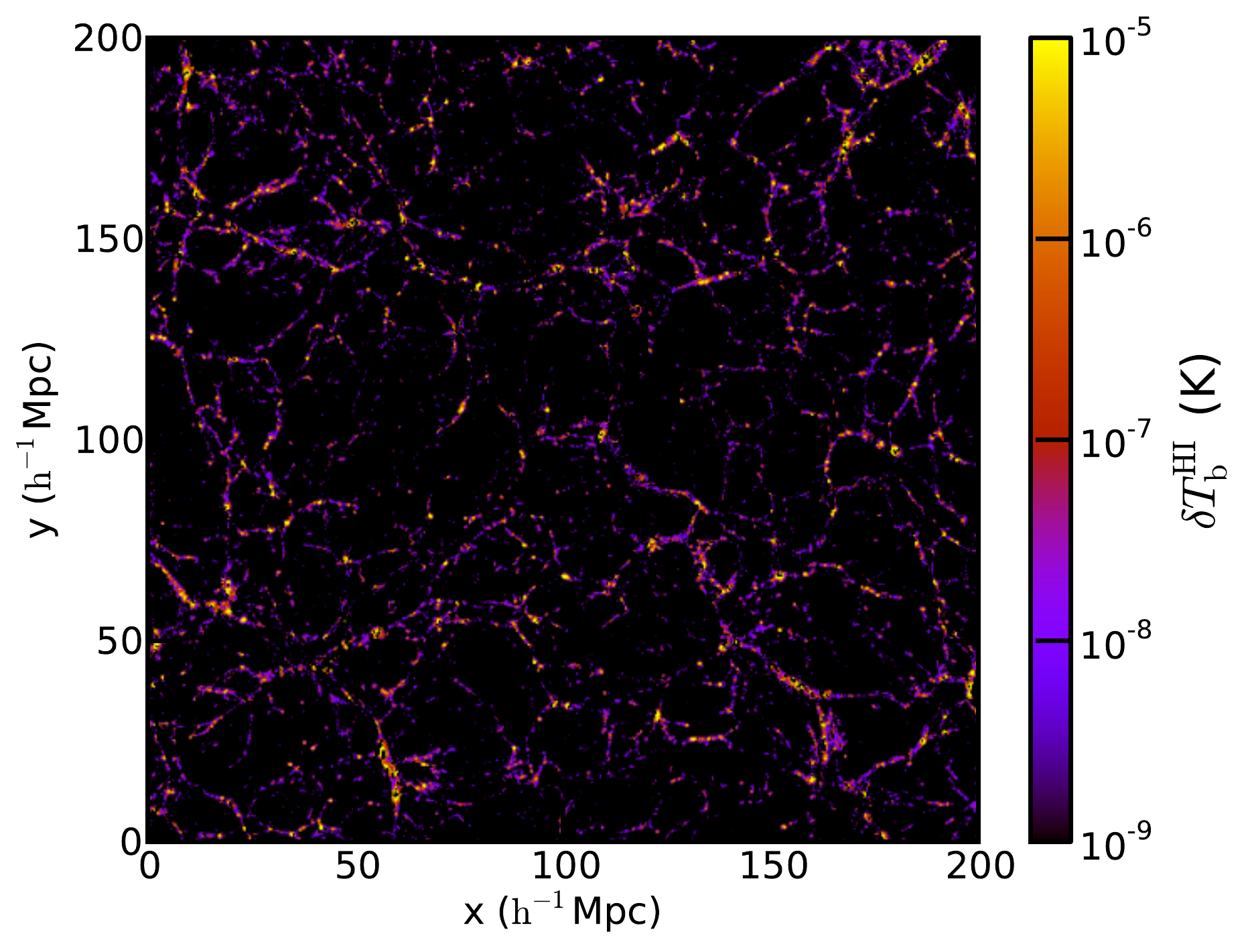}
\caption{Slice of the simulation box at z = 0.1 with a width of 0.33 $h^{-1}\,$Mpc. The left panel shows the overdensity $\delta$ and the right panel shows the differential brightness temperature of HI. Cells with overdensities $\Delta_b \geq \Delta_c$ have been masked in the differential brightness temperature box.}
\label{fig:boxes}
\end{centering}
\end{figure*} 

\section{Cosmological simulations}\label{sec:sim}
In order to predict the signal from more realistic filaments we make use of high resolution simulations which 
make it possible to resolve different filament morphologies and allow for differences in the gas properties along a filament.

We start by running the parallel Tree-Particle Mesh code Gadget 2 \citep{art:gadget2} with $1024^3$ 
particles and a volume of $\left(200\,h^{-1}\mathrm{Mpc}\right)^3$. This corresponds to a mass resolution of 
$6.514\times10^8\, \mathrm{M_\odot} h^{-1}$. 

We then use cloud-in-cell interpolation to divide the simulation particles into a grid of N = $600^3$ cells, 
corresponding to a spatial resolution of $\sim0.33\, {h^{-1}\, {\rm Mpc}}$. This is enough to properly resolve the filament morphology.
Then we determine the density contrast $\delta = \rho/<\rho>-1$ for each cell of the simulation and, assuming that
the spatial distribution of the baryons follows that of the dark matter, we compute the gas temperature, ionization state and 
differential brightness temperature following the prescriptions described in Sections \ref{secmodel} and \ref{sec:sig}. 

Because we are interested in the emission of the IGM and not from galaxies, we 
mask the cells that are above the virialized limit determined through the scaling relations by \citet{art:collapse}. 
For z = 0.1 this critical overdensity is $\Delta_c \approx 297$. Figure \ref{fig:boxes} shows a slice of the simulation at 
z = 0.1, with a thickness of 0.33 $h^{-1}$Mpc, where several large filaments can be visually identified.

Additionally, we constructed a catalogue with the positions and masses of dark matter haloes using the Amiga halo finder 
\citep{art:halofind} on the initial particle catalogue. At low redshift, the most massive haloes are expected to have more 
than one galaxy. However most of the luminosity of a halo usually originates in a single bright galaxy. This is also the 
only galaxy that a modest galaxy survey would probably detect, so it is reasonable to assume that each of these haloes 
would correspond to a single galaxy when comparing with SDSS detected galaxies. 

\section{21cm emission from SDSS Filaments}\label{sec:fils}
In this section we select gas filaments indirectly detected through the SDSS galaxies and estimate their integrated HI intensity. 
Given that galaxy surveys, such as the SDSS, do not probe the gas in filaments, we  
estimate the properties of the filamentary gas from similar length filaments found in our cosmological N-body simulations, which 
contain a similar number of galaxies. 
We start with a description of 
the catalogue and the criteria that were used to select the filaments and follow with the estimation of the HI 21 cm line intensity. 
The estimates are then used to determine the detectability of these filaments by current and future telescopes.

\subsection{Filament catalogue}
Since we are interested in targeting filaments inferred from galaxy catalogues, a complete sample of galaxies, covering a 
relatively large volume and  down to a relatively low magnitude threshold is required to identify large scale filaments, the 
largest of which, currently available was obtained by SDSS. 
The sample includes 499340 galaxies and goes up to a redshift of $z$ = 0.155 with a lower limit of 
$z$ = 0.009. The lower magnitude limit of the sample of galaxies is set to $m_{\mathrm{r}}$ = 17.77, imposed by the limits of the 
spectroscopic sample \citep{art:ssdssspecsamp}. 

\begin{figure} 
\begin{centering}  
\includegraphics[angle=0,width=0.45\textwidth]{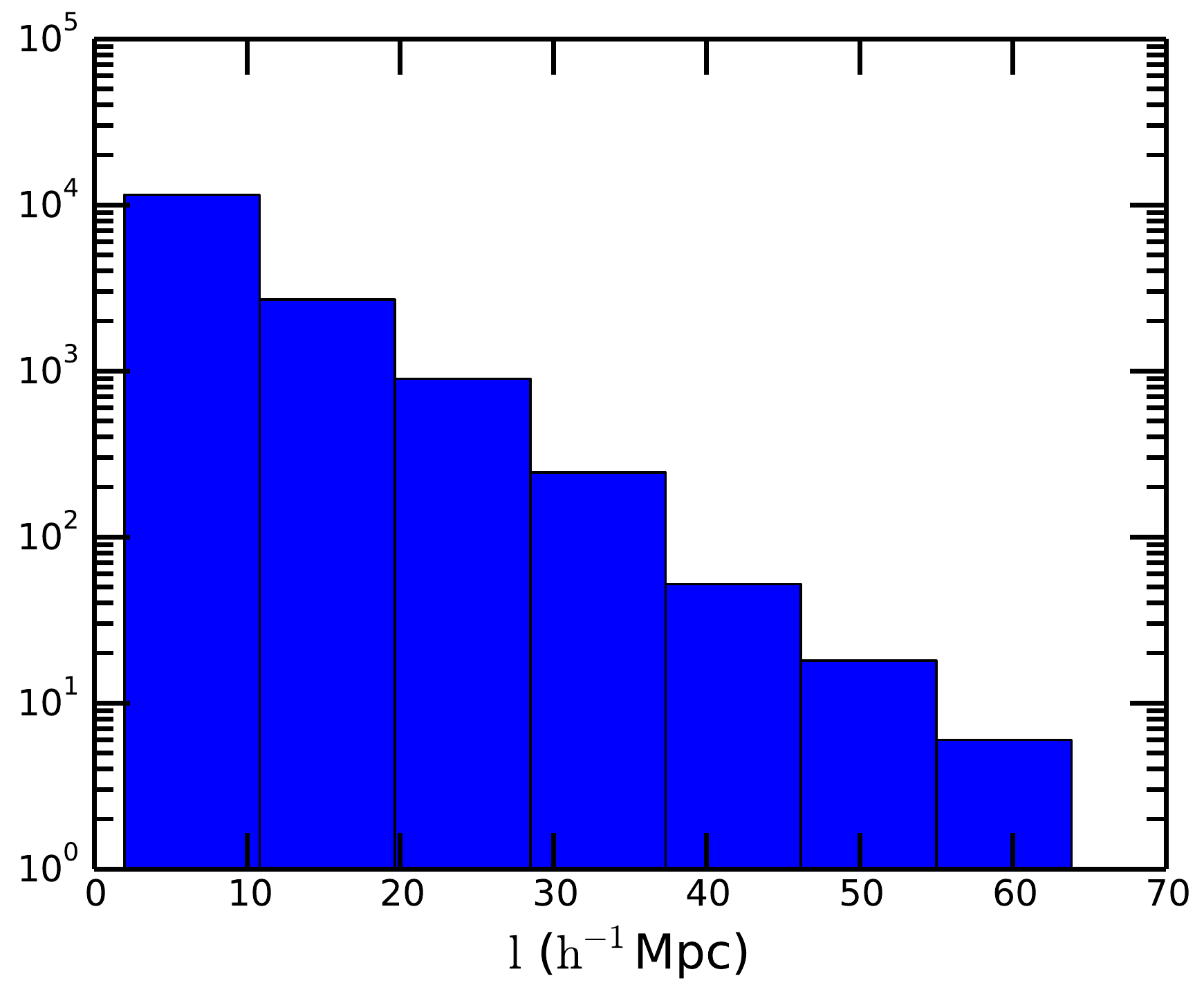}
\caption{Distribution of the length of the filaments found in the \citet{art:tempel} catalogue.}
\label{fig:fildist}
\end{centering}
\end{figure}

We therefore use the filament catalogue by \citet{art:tempel}. 
This catalogue is obtained by statistically inferring the filamentary pattern in the SDSS data release 8 sample through the 
Bisous model. The model assumes a fixed maximum distance to which galaxies can be separated from the filament spine and still belong to it, 
which in this case was chosen to be 1 $h^{-1}$Mpc. 
In Figure \ref{fig:fildist} we show the length distribution of the filaments from the catalogue. Most filaments are short 
(less than 10 $h^{-1}$Mpc) and would therefore result in a low signal-to-noise in most observations. However, a sample of $\sim$ 
4000 long filaments, reaching lengths of $\sim$ 10 $h^{-1}$Mpc to $\sim$ 50 $h^{-1}$Mpc is also found. Within this sample of longer 
filaments we choose the most suitable candidates to be targeted by observations, taking into account not only their lengths, but also their galaxy densities. It should be noted here that, given the spatial density of galaxies in the SDSS catalogue, the \citet{art:tempel} method  tends  to fragment the filaments and assign smaller lengths to them than in reality. 

The signal from a filament is highest when it can be observed directly along the line of sight, 
because then a larger part of the filament can be integrated over with a single pointing of the 
telescope. Therefore, we select the larger filaments with the smaller alignment angles ($\theta$) 
with the line of sight. For simplicity we define the inclination by assuming a straight line 
between the endpoints of the filaments. This method might exclude long filaments whose endpoints 
are not aligned along the line of sight, but that in between do have large parts that are aligned. 
However, we still find a number of useful filaments. In Table \ref{tab:filselect} we give the properties of the seven selected filaments.
\begin{table} 
\caption{Properties of SDSS filaments with small alignment angles. The parameter $l$ is the length of 
the filament, $\theta$ is the alignment angle and $N_{\mathrm{gal}}$ is the number of galaxies associated with the filament.}
\begin{center}
\begin{tabular}{l|c|c|c|c|c}
 \hline
 \hline
 ID & z & $d_{\mathrm{com}}$ ($h^{-1}$Mpc) & $l$ ($h^{-1}$Mpc) & $\theta\, (^\circ)$ & $N_{\mathrm{gal}}$\\
 \hline
 1 & 0.05 & 175 & 23.5 & 5.02 & 44\\
 2 & 0.04 & 130 & 13.9 & 5.70 & 32\\
 3 & 0.07 & 227 & 16.3 & 4.27 & 40\\
 4 & 0.11 & 333 & 16.8 & 1.02 & 44\\
 5 & 0.12 & 356 & 14.1 & 2.05 & 55\\
 6 & 0.10 & 282 & 16.6 & 2.21 & 42\\
 7 & 0.06 & 180 & 19.2 & 5.42 & 21\\
 \hline
\end{tabular}
\end{center}
\label{tab:filselect}
\end{table}
Filaments 4, 5 and 6 are the most aligned along the line of sight. They have lengths of $l\sim$ 15 $h^{-1}$Mpc, are 
all relatively straight spatially and are therefore ideal candidates for observations. 

\subsection{Signal estimation}

Ideally one would like to directly link current observables (galaxies) to the gas density in the filaments. 
Some work has been done to find a relation between the luminosity density of galaxies and the warm hot intergalactic 
medium (WHIM) using simulations \citep{art:nevalainen}. However, there are no similar relations between galaxy luminosity 
and the properties of the cold gas in filaments. 
Here we do a similar exercise in combining observations with simulations by looking for similar 
filaments to the ones found by SDSS. From the simulated box we visually identify a number of long filaments 
which, after a careful selection, can be used to estimate the expected HI signal of the SDSS filaments.
Although most of the selected filaments, in the \citet{art:tempel} catalogue, have lengths of $\sim$ 15 $h^{-1}$Mpc, the real gas filaments can be 
more extended, or the catalogue model can identify smaller parts of a larger filament as separate ones. Consequently, the observed signal 
can be higher than what would be expected by assuming the length inferred from the galaxy distribution. Therefore, in the simulations we 
search for filaments that are longer ($\sim$ 50 $h^{-1}$Mpc, similar to the maximum length found in the filament catalogue) and, by comparing 
to the haloes in the simulation, determine what part of the filament would be detected by the SDSS.

Although, our simulation box is at a redshift of $z$ = 0.1, the relevant properties of the gas in the filaments 
is not expected to vary by much for the redshift range covered by the SDSS selected filaments.

\begin{figure} 
\begin{centering}  
\includegraphics[angle=0,width=0.48\textwidth]{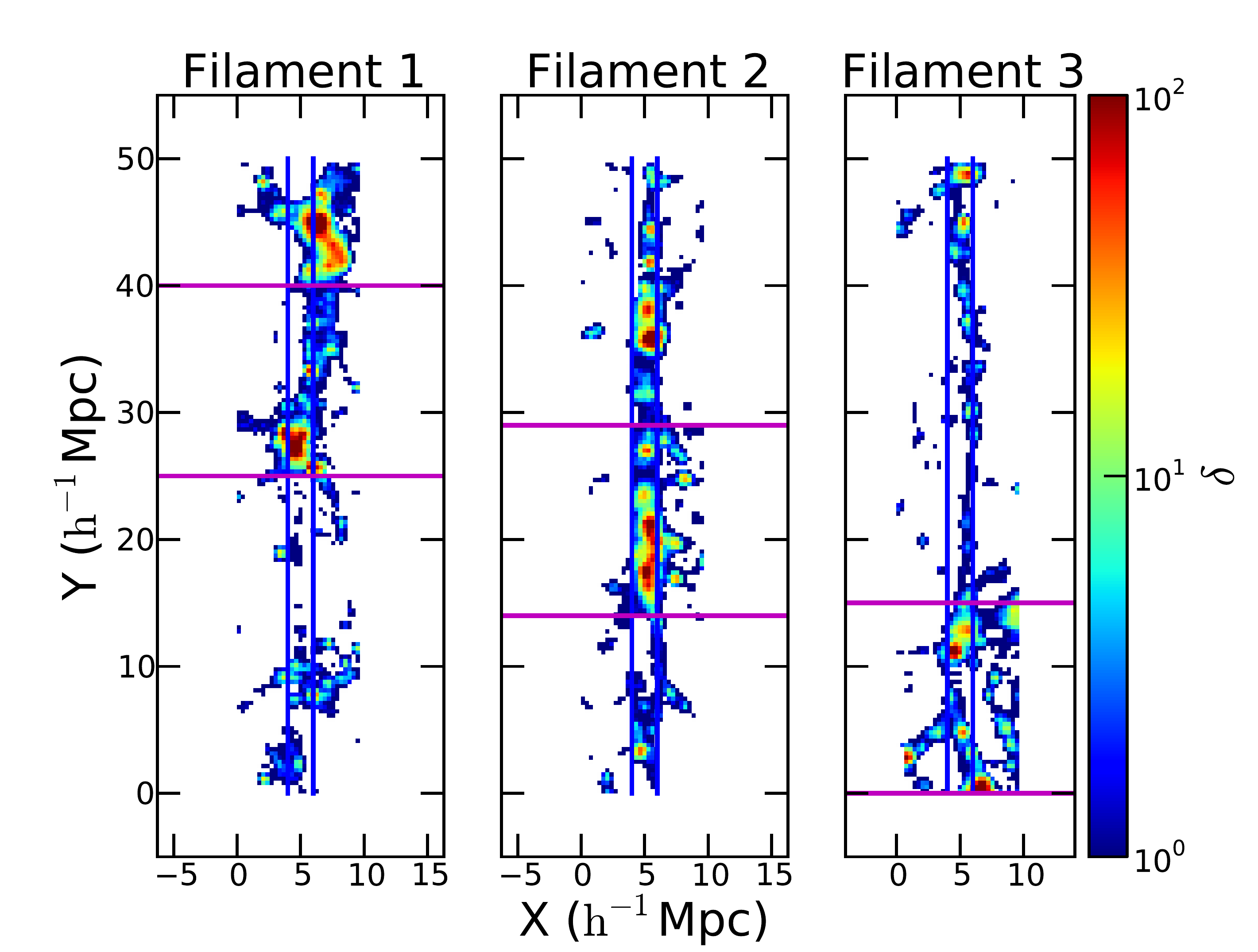}
\caption{The density fields of the selection of filaments from the simulation. The blue lines denote 
the observational skewer, whereas the magenta line shows what part would be preferentially detected by SDSS. The colourbar denotes 
the mean overdensity over the slices in the z-direction.}
\label{fig:filsel}
\end{centering}
\end{figure}

Figure \ref{fig:filsel} shows the extracted filaments from the simulation box. They each have a length 
of $\sim$ 50 $h^{-1}$Mpc and a radius of $\sim$ 1 $h^{-1}$Mpc and are thus comparable to the filaments 
that can be detected by SDSS. 
We now analyse in more detail three filaments that cover the diversity of the selected filaments 
and the expected range in their intensities.
Filament 2 is relatively straight along the line of sight, compared to the 
other two and is therefore the optimistic case. Filament 3, on the other hand, is relatively faint at the 
top and middle sections and therefore most of the signal arises from the bottom part of the filament as shown in Figure \ref{fig:filsel}. 
This is our pessimistic case. Filament 1 is somewhere in between the other two in terms of their 
expected signal strengths.

The signal of the simulated filaments is determined by taking a cylindrical skewer centered on the filament, 
which represents the beam of the telescope and then the signal of the filament is calculated as follows
\begin{equation}
 \delta T_{\rm b}^{\rm fil} = \frac{\sum_{i,j}\delta T_{\rm b}^{i,j}}{\Delta R\cdot\pi r_{\mathrm{S}}^2},
\end{equation}
where $\delta T_{\rm b}^{i,j}$ is the signal per cell within the skewer, $\Delta R$ is the length of the skewer, 
which depends on the observed bandwidth, and $r_\mathrm{S}$ is the radius of the skewer, which is determined by the 
angular resolution. We set $\Delta R$ = 50 $h^{-1}$Mpc (corresponding to a frequency bandwidth of $\sim$ 15 MHz) and 
the radius to $r_\mathrm{S}$ = 1 $h^{-1}$Mpc (or an angular resolution of 10 arcmin). The blue lines in Figure \ref{fig:filsel} 
indicate the observational skewer. By comparing to the simulated halo map previously obtained with the Amiga halo finder, we 
find that most of the massive haloes are located in the region between the magenta lines and a galaxy survey like SDSS 
would therefore most likely observe the regions between these lines.

\subsection{Observability}\label{sec:observability}
\begin{table*} 
\caption{Telescope parameters. Arecibo and FAST are single dish instruments, whereas the others are interferometers. The parameter $D_{\mathrm{dish}}$ Dish diameter, $N_{\mathrm{dish}}$ denotes the number of dishes, $A_{\mathrm{tot}}$ is the total (illuminated) surface area, $\epsilon_{\mathrm{ap}}$ gives the aperture efficiency, $T_{\mathrm{sys}}$ is the system temperature for the observed frequency band, $D_{\mathrm{max}}$ the maximum baseline length, $\theta_{\mathrm{res}}$ the angular resolution and $\nu_{\mathrm{res}}$ is the minimum possible frequency resolution.}
\label{tab:tels}
\centering
 \begin{tabular}{l|c|c|c|c|c|c|c|c|c|c}
 \hline
 \hline
 Telescope & $D_{\mathrm{dish}}$ (m) & $N_{\mathrm{dish}}$ & $A_{\mathrm{tot}}$ ($\mathrm{m}^2$) & $\epsilon_{\mathrm{ap}}$ & $T_{\rm sys}$(K) & Spectral range (GHz) & $D_{\mathrm{max}}$ (km) & $\theta_{\mathrm{res}}$ (') & $\nu_{\mathrm{res}}$ (kHz) & FoV ($\mathrm{deg^2}$)\\
 \hline
 Arecibo & 205 & - & 32,750 & 0.7 & 30 & 0.047 - 10 & 0.3 & 3.24 & 12.2 & 0.17\\
 FAST & 300 & - & 70,700 & 0.55 & 25 & 0.070 - 3 & 0.5 & 2.9 & $\lesssim$ 0.5 & 0.36 \\
 Apertif (WSRT) & 25 & 12 & 5,890 & 0.75 & 55 & 1.13 - 1.75 & 2.7 & 0.36 & 12.2 & 8 \\
 EVLA & 25 & 27 & 13,300 & 0.45 & 26 & 1 - 50 & 1 - 36 & 0.97 - 0.03 & 31 & 0.42\\
 GMRT & 45 & 30 & 47,720 & 0.4 & 75 & 0.05 - 1.5 & 25 & 0.04 & 31 & 0.13\\
 ASKAP & 12 & 36 & 4,072 & 0.8 & 50 & 0.7 - 1.8 & 6 & 0.5 & 18.3 & 30\\
 MeerKAT & 13.5 & 64 & 9,160 & 0.8 & 20 & 0.580 - 14.5 & 20 & 0.05 & $\lesssim$ 18 & 1.44 \\
 SKA-2 & 15 & 1,500 & 300,000 & 0.8 & 30 & 0.070 - 10 & 5 (core) & 0.19 & $\lesssim$ 18 & 1.17\\
 \hline
 \end{tabular} 
\end{table*}

\begin{figure*} 
\begin{centering}
\includegraphics[angle=0,width=\textwidth, height=300 pt]{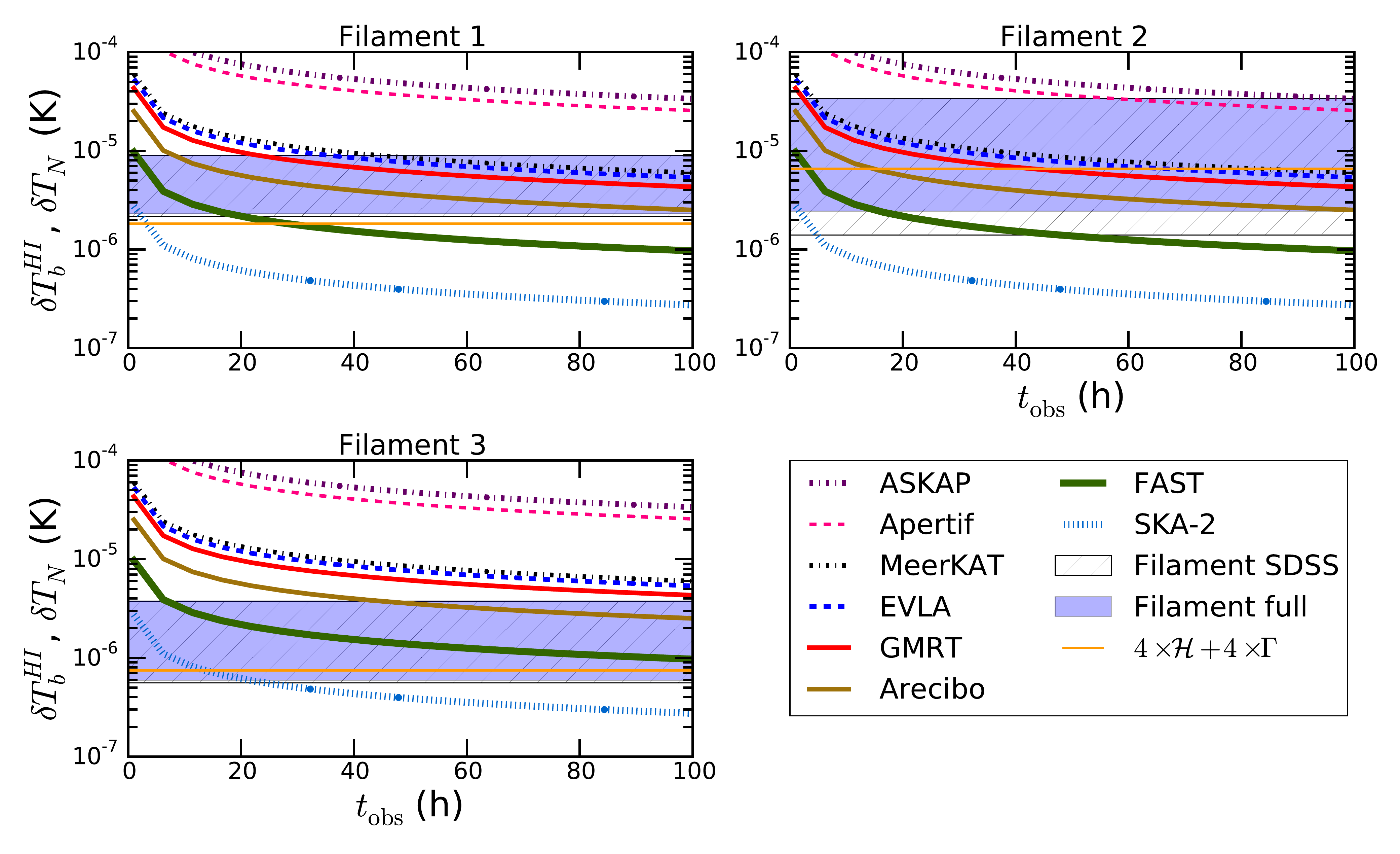}
\caption{The expected signal of the three filaments in this study (see Figure \ref{fig:filsel}) together with 
the noise temperatures for the different instruments being considered. The shaded blue area shows the signal for the full filament, 
where the top and bottom denote the minimum and maximum signal when rotating the observational skewer -5 to +5 degrees. 
The white striated shaded area shows the same, but for the case where the filament is only as long as expected from SDSS data.  
The coloured lines denote the noise level of the instruments described in Table \ref{tab:tels} for $\Delta\theta$ = 10 arcmin,
$\Delta\nu$ = 15 MHz. The orange solid line shows the maximum signal of the filament after increasing the heating and the photoionization by a factor of 4.}
\label{fig:filsig}
\end{centering}
\end{figure*} 

In order to check the observability of these filaments, we compare the signal we obtain with the sensitivity of multiple instruments for the same conditions. 
In general, the noise in the measurement of a radio telescope can be written as  \citep{art:furlanetto06} 
\begin{equation}
\delta T_{\rm N} = \frac{c^2(1+z)^2}{\nu_{0}^2\Delta\theta^2\epsilon_{\rm ap}A_{\rm dish}}\frac{T_{\rm sys}}{\sqrt{2\Delta\nu t_{\rm obs}}},
\end{equation}
where $\epsilon_{\mathrm{ap}}$ is the aperture efficiency, $A_{\mathrm{dish}}$ the total (illuminated) surface area of a single dish of the array, $\Delta\theta$ the size of the beam, $\Delta\nu$ the frequency bandwidth and $t_{\mathrm{obs}}$ the observation time. The factor 2 in the last term follows from observing two polarizations simultaneously and integrating them together. 
The system temperature $T_{\mathrm{sys}}$ of a radio telescope has two components, one due to the sky that dominates at low frequencies and another due to the receiver, dominant at high frequencies. The brightness temperature uncertainty (sensitivity) $\delta T_{\mathrm{N}}$ is thus also a combination of the two contributions, where in the case of an interferometer the noise drops by the square root of the number of baselines ($N_{\mathrm{B}}=N_{\mathrm{dish}}(N_{\mathrm{dish}}-1)/2$), giving 
\begin{equation}
 \delta T_{\mathrm{N}} = \left(\delta T_{\mathrm{N}}^{\mathrm{sky}} + \delta T_{\mathrm{N}}^{\mathrm{rec}}\right) \times \begin{cases}
          1 & (\mathrm{single\, dish})\\
          1/\sqrt{N_{\mathrm{B}}} & (\mathrm{interferometer})
          \end{cases}
          .
\end{equation}

We note that this calculation assumes that the filaments contain structure on all scales for which the interferometers have baselines and therefore do not suffer from spatial filtering. This will be discussed in more detail in Kooistra et al. in prep.\\

Ideal instruments would be those that have both a large field of view and good sensitivity in order to be able to probe the extended low-surface 
brightness HI emission. We consider both single dish telescopes and interferometers. The single dish telescopes are Arecibo and FAST. For interferometers 
we consider Apertif on the Westerbork Synthesis Radio Telescope (WSRT), the Expanded Very Large Array (EVLA), GMRT, the Australian Square Kilometre Array 
Pathfinder (ASKAP), the Karoo Array Telescope (MeerKAT) and the second phase of the Square Kilometre Array (SKA-2). The relevant parameters for each of 
them are given in Table \ref{tab:tels}.

In Figure \ref{fig:filsig} we show the brightness temperature uncertainties for each of the instruments as a function of the observation time. Each plot 
gives a comparison with the signal of one of the filaments, previously shown in Figure \ref{fig:filsel}, where the blue shaded region gives the signal 
of the full simulated filament and the white striated region denotes the signal from the filamentary region that we expect to be detected from SDSS data. 
The minima and maxima of these regions denote the minimum and maximum signal that arises when rotating the observational skewer from -5 to +5 degrees 
with respect to the filament spine. This uncertainty accounts for a possible misalignment of the telescope beam and the filament spine, 
which we need to account for, given that this spine has to be defined without information on the gas content of the filament.

Depending on the morphology of the filament, the small rotation of $\pm$ 5 degrees can change the signal of the filament 
by more than an order of magnitude. Since the regions corresponding to the SDSS filaments also contain the 
brightest parts of the filaments, the difference between the signal from either just that region or the full filaments is usually small. 
As can be clearly seen in Figure \ref{fig:filsig}, a number of instruments should be able to observe the HI signal within $\sim 100$ hours. The SKA will be able to detect the signal in all cases. Furthermore, single dish telescopes are the best alternatives, where FAST can detect the signal in all but the worst case scenario. The signal is also detectable for most instruments in the most optimistic case, whereas for the lower signals from Filaments 1 and 3, the signal would still be within reach of FAST, Arecibo and the SKA. Apertif and ASKAP can only make a low signal-to-noise detection of the strongest filament signal considered here.\\
There is some uncertainty in the amount of heating that goes into the IGM. Since we do not include a prescription for shock heating in this medium, we might be underestimating the amount of heating and as a consequence also underestimate the ionization fractions. The amount of shock heating predicted by simulations differs a lot, depending on the assumptions on the feedback and on how it is implemented. According to the EAGLE simulation \citep{art:eagle} this is, however, not a crucial effect in the gas we target and so most of the gas in the EAGLE simulation follows the same temperature density relation we use in our simulation \citep{art:crain17}. Nevertheless, we account for an extra source of heating by manually increasing the heating and photoionization rates of the \citet{art:hm12} background by a factor of 4. The resulting signal from the filaments reduces by a factor of $\sim$5, as shown by the orange lines denoting the maximum signal under these conditions in Figure \ref{fig:filsig}. Even in this pessimistic case the maximum signal from the filaments would be detectable by the SKA. For strong filaments, such as filament 2, it would be possible to get a good detection with some of the other instruments as well.

\section{The Apertif and ASKAP HI surveys}\label{sec:surveys}
In the previous sections we studied the detectability of the HI gas in IGM filaments for the case where the filaments are aligned along the line of sight and so their signal can be integrated over a single pointing, even by instruments with a small FOV. However, some of the considered instruments 
(i.e. Apertif, ASKAP and SKA) have a very large field of view and so they can, with a single pointing, also detect filaments with different alignments or even curved filaments. 
Furthermore, Apertif and ASKAP have a number of planned deep HI surveys in the near future. 

For Apertif there 
will be a medium deep survey\footnote{http://www.astron.nl/radio-observatory/apertif-surveys} covering 450 $\mathrm{deg}^2$ with 84 hours integration for each pointing. 
ASKAP will have the Deep Investigation of Neutral Gas Origins (DINGO) survey\footnote{http://askap.org/dingo}. DINGO will cover two fields on the southern hemisphere, smaller 
but with longer integration times than the full ASKAP field: the DINGO-Ultradeep field of 60 $\mathrm{deg}^2$ with pointings of 2500 
hours and the less deep, but larger DINGO-Deep field of 150 $\mathrm{deg}^2$ and 500 hour pointings. 
The areas covered by these surveys are larger than the ones in our simulation and so they are likely to contain filaments at least as bright in HI as the ones considered here. 

In this section we present the feasibility of detecting curved filaments by such surveys. We assume that the spatial location of the filament can be inferred
through the positions of the bright galaxies previously observed by the same instrument or by other instruments. In this case it is possible to follow the spine 
of the filament for the integration. 

\begin{figure} 
\begin{centering}
\includegraphics[angle=0,width=0.49\textwidth]{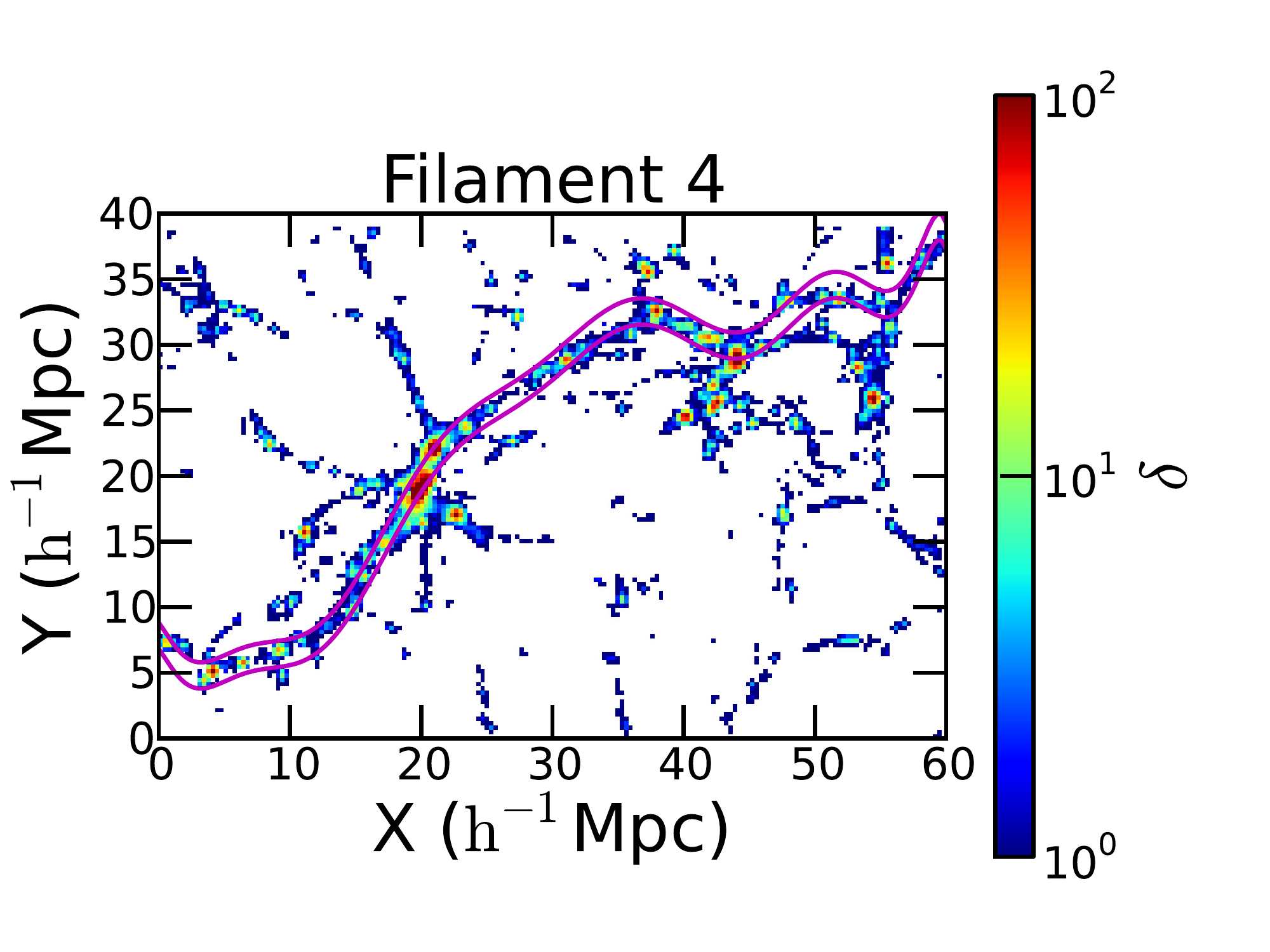}
\caption{The density field of the selected curved filament from the simulation. The magenta lines show
the observational skewer over which the integration was performed with an angular resolution of 10 arcmin and a frequency bandwidth of 
0.6 MHz, corresponding to a filament radius of $\sim$ 1 $h^{-1}\mathrm{Mpc}$. The colourbar denotes 
the mean overdensity over the filament in the z-direction.}
\label{fig:filselcurv}
\end{centering}
\end{figure}

\begin{figure} 
\begin{centering}
\includegraphics[angle=0,width=0.48\textwidth]{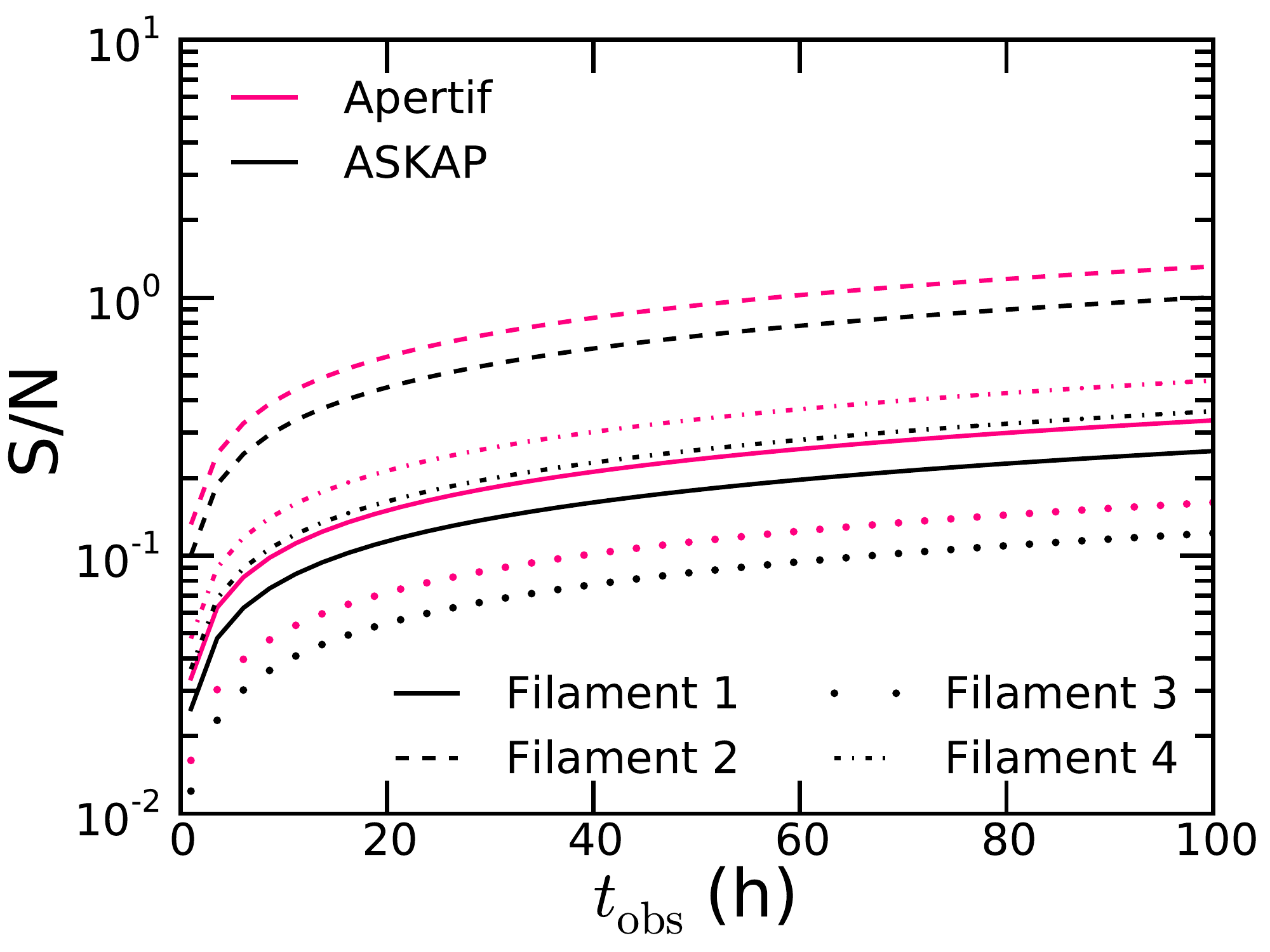}
\caption{Expected signal to noise of the simulated filaments with the HI survey instruments Apertif and ASKAP. We assume an angular resolution of 10 arcmin and a frequency bandwith of 0.6 MHz. The colour of the lines denotes the instrument and the linestyle shows for which filament it is.}
\label{fig:filsigsurv}
\end{centering}
\end{figure}

We extract a curved filament from our simulation, as can be seen in Figure \ref{fig:filselcurv} and estimate its detectability.
Filament 4 is similar in density and length as the previously discussed aligned filaments. In order to calculate a curved filament signal, we consider that each small section of the filament is aligned perpendicular to the line of sight and that the instantaneous field of view of the survey encompasses the entire filament. This is a reasonable assumption for the case of both Apertif and ASKAP observations at $z$ = 0.1. Then we integrate along the filament by again assuming an angular resolution element of 10 arcmin, but now the frequency bandwidth will only be 0.6 MHz for the part along the line of sight (in the z-direction in Figure \ref{fig:filselcurv}). The section over which we integrate the filament is shown by the magenta lines in Figure \ref{fig:filselcurv}.\\ 
In this case the noise of the telescope for the integration of a single segment of the filament will be higher than for the case of the aligned filaments 
discussed in Section~\ref{sec:sim}. Fortunately, the integration is done over multiple of these 10 arcmin resolution segments and the noise drops as the square root of this number. 
The resulting signal to noise of the four filaments for both telescopes is shown in Figure \ref{fig:filsigsurv}. Filament 2 is the brightest and is thus the only one that can be detected by both Apertif and ASKAP in 100 h. The other filaments fall below the detection threshold and we note that when the heating and photoionization rates are increased, the signal will become hard to detect with these surveys. In all cases, the signal-to-noise is almost the same as when the filament was aligned along the line of sight, showing that the orientation of a filament matters little if the FoV of the survey is large enough to cover the entire spine. 

With their wide fields of view, survey instruments like Apertif and ASKAP might thus provide excellent tools for detecting HI emission from IGM filaments, provided that the HI gas in these filaments correlates well with the positions of the strongest galaxies in the same field. 
The cross-correlation of the HI signal with the galaxies could also possibly be used to confirm wether the signal really corresponds to a filament. Furthermore, since the noise from both quantities is not correlated, to first order, the noise terms would cancel in the cross-correlation, which could render the cross-correlation itself another useful probe of the large scale filaments to be used for the understanding of galaxy formation and evolution mechanisms. This will be explored further in a future publication.

\section{Contamination from galaxies}\label{sec:contam}
Besides the IGM, another source of HI 21 cm emission in filaments are the galaxies. These are brighter than the IGM itself 
and will thus contaminate the integrated signal. Fortunately, the HI signal from galaxies is largely dominated by a 
few bright galaxies, which can be directly observed by current instruments, such as  Arecibo or WSRT, and certainly by SKA-2. 
Also, these galaxies are relatively small compared to the width of the filament, given that typical sizes of HI disks are of the order 
of $\sim$30-60 kpc \citep[e.g.,][]{art:h1sizegals}. In the above sections we have calculated the signals by masking all cells 
in our simulation box with a density above the critical density for collapse. This corresponds to 0.4$\%$, 0.6$\%$, 0.1$\%$ and 0.4$\%$ 
of the pixels in the skewer for the proposed pixel size of $(0.33\,h^{-1}\mathrm{Mpc})^3$ for filament 1, 2, 3 and 4, respectively. We note that the resolution available from
the proposed instruments is usually even higher. 
This means that, assuming that all dark matter haloes contain galaxies, the emission from all galaxies in the filament could be masked, 
without losing much of the IGM filamentary emission. In reality it is only necessary to know the positions of the brightest galaxies and mask the 
corresponding pixels in the observation. This requires deep galaxy surveys and instruments that have high resolution, both 
spatially and in frequency. The instruments that were considered in Section \ref{sec:observability} all have the necessary resolution to 
be able to mask the most luminous foreground galaxies, without erasing the target signal. In most cases, these surveys can also detect the 
position and the brightness of these galaxies. In the case of the filaments detected by SDSS, the positions of the galaxies are already 
known from the survey itself and they only need to be masked.

To check the level of contamination to the signal by galaxies, we make a conservative upper estimate of how much HI would be contained 
by them and compare it to the HI content of the IGM in the filaments in the simulation. The halo finder provides the masses of the dark 
matter haloes. We can convert these masses to HI masses by using the fitting function to the HI mass-halo mass relation determined through 
abundance matching by \citet{art:haloh1}:
\begin{align}
 M_{\mathrm{HI}} = & 1.978\times10^{-2}M_{\mathrm{h}}\times\notag\\
 &\left[\left(\frac{M_{\mathrm{h}}}{4.58\times10^{11}\,\mathrm{M_\odot}}\right)^{-0.90}+\left(\frac{M_{\mathrm{h}}}{4.58\times10^{11}\,\mathrm{M_\odot}}\right)^{0.74}\right]^{-1}\label{HIHM}.
\end{align}
$M_{\mathrm{h}}$ denotes the mass of the halo and $M_{\mathrm{HI}}$ gives the mass of the HI expected to be inside the halo. The total HI mass in galaxies in the simulated filaments are $2.8\times10^{11}\,\mathrm{M_\odot}$, $5.5\times10^{11}\,\mathrm{M_\odot}$, $2.1\times10^{11}\,\mathrm{M_\odot}$ and $4.0\times10^{11}\,\mathrm{M_\odot}$ for filament 1, 2, 3 and 4, respectively. Their total HI masses contained by the IGM are, respectively, $3.5\times10^{12}\, \mathrm{M_{\odot}}$, $5.2\times10^{12}\, \mathrm{M_{\odot}}$, $1.7\times10^{12}\, \mathrm{M_{\odot}}$ and $4.7\times10^{12}\, \mathrm{M_{\odot}}$.

Galaxy surveys can only find the locations of the brightest galaxies. The flux limit of the survey constrains which fraction of the  contaminating galaxies can be masked from observations. In order to get an estimate of the amount of HI mass that could be inferred from SDSS data, we convert the flux limit of SDSS ($m_\mathrm{r}^{\mathrm{lim}}$ = 17.77) to a minimum galaxy HI mass ($M_{\mathrm{HI}}^{\mathrm{min}}$) in three steps. We connect the SDSS magnitude limit to the typical colour of galaxies using the colour-magnitude diagram of \citet{art:mrtocolor}. In this case the highest limit (thus the least effective in removing the contamination by galaxies) comes from assuming a blue galaxy, giving a colour of $(u-r)_{0.1} \approx$  1.85. Using this colour we then find the appropriate mass-to-light ratio for these galaxies from \citet{art:colortomasslight}, corresponding to $\log\left(M_*/L\right) \approx$ 0.33, which, for the luminosity limit of SDSS, corresponds to a stellar mass of $\log M_* \approx$ 9.62. Finally, we convert the stellar mass into an HI mass by applying the stellar mass to gas mass relation found by \citet{art:stellartogasmass}, which results in a limiting HI mass for SDSS of $M_{\mathrm{HI}}^{\mathrm{lim}} \approx 2\times10^{9} \mathrm{M_{\odot}}$. By masking all the haloes in our simulated filaments with HI masses above this threshold, we then estimate the remaining contamination.
\begin{table} 
\caption{HI masses of the simulated filaments. $M_{\mathrm{HI}}^{\mathrm{IGM}}$ is the HI mass in the IGM of the filament, $M_{\mathrm{HI}}^{\mathrm{gal}}$ gives the 
total HI mass contained by the haloes in the filaments and $M_{\mathrm{HI}}^{\mathrm{cut}}$ gives the remaining HI contamination after removing the haloes with masses above the SDSS flux limit. All masses are given in $\mathrm{M_{\odot}}$.}
\centering
\begin{tabular}{c|c|c|c}
 \hline
 \hline
 Filament & $M_{\mathrm{HI}}^{\mathrm{IGM}}$ & $M_{\mathrm{HI}}^{\mathrm{gal}}$ & $M_{\mathrm{HI}}^{\mathrm{cut}}$\\
 \hline
 1 & $3.5\times10^{12}$ & $2.8\times10^{11}$ & $4.5\times10^{10}$\\
 2 & $5.2\times10^{12}$ & $5.5\times10^{11}$ & $5.9\times10^{10}$\\
 3 & $1.7\times10^{12}$ & $2.1\times10^{11}$ & $4.3\times10^{10}$\\
 4 & $4.7\times10^{12}$ & $4.0\times10^{11}$ & $5.1\times10^{10}$\\
 \hline
\end{tabular}
\label{tab:HIcont}
\end{table}
Table \ref{tab:HIcont} gives a summary of the results for each of the four filaments. As can be seen, the remaining contamination due to galaxies is two orders of magnitude smaller than the HI content in the IGM of the filaments. A survey like SDSS will thus be sufficient for dealing with the most severe contamination due to galaxies.

\section{Conclusions}\label{sec:concl}
In this work we explored the possibility of observing the integrated HI 21 cm line emission from large scale filaments in the IGM. 
Directly mapping this emission is a potential new avenue to probe the spatial distribution of the filaments, and their gas content, ionization state and temperature. 
The properties of the filaments can be useful to construct more realistic models of galaxy formation and evolution, 
given the constant exchange of gas between galaxies and their surrounding medium, which is usually a large scale IGM 
filament. Moreover, since the thermal and ionization state of the gas far from strong local sources is dominated by the UV background, these properties can be 
used to constrain this radiation field.

Our study has mainly focused on filaments at z=0.1, primarily because that is the average 
redshift at which good galaxy catalogues are available. The new generation of surveys will, however, also provide good galaxy catalogues at higher redshifts, and the proposed SKA-2 survey can be used to probe their emission. This would make it possible to constrain the evolution of the UV background even further.

In this study, we took observed filaments inferred from SDSS data and estimate their integrated HI 21 cm signal and its detectability by current and upcoming surveys.
By selecting the largest filaments with the smallest inclination to the line of sight, we determined the intensity of the easier to observe filaments to be of the order of $10^{-6}-10^{-5}$K.
We check the observability of these signals for an integration time of up to 100 hours and found that a number of the radio surveys considered could  detect some of these filaments in less than 50 hours.
In particular, FAST and the SKA are good candidates to detect the signal.

The upcoming HI surveys for instruments with large fields of view, such as Apertif and ASKAP, could furthermore remove the need for observing spatially straight filaments aligned along the line of sight. We find that the integration times of the planned surveys for these two instruments will be sufficient to make the detection of the IGM gas in the strongest filaments feasible and opening up an interesting avenue to explore with these instruments.

Contamination to the signal from galaxies has to be taken into account, given that the bulk luminosity at the 21cm line emanates from galaxies and not from the IGM gas we are targeting. 
We used the observational luminosities of the galaxies obtained by SDSS and compared them to simulated luminosities. 
This allowed us to safely conclude that this emission is dominated by a small number of sources, whose position can be 
determined by galaxy surveys and masked from observations before the signal is integrated. We note that these filaments 
inferred from SDSS data are biased towards the most luminous galaxies and should therefore have considerably more galaxy 
contamination than most IGM filaments. Our estimates also show that this masking procedure would bring the contamination 
down to a negligible level compared to the total signal of the filament, without erasing the target signal, even when 
attributing more HI mass to the remaining galaxies than what they are expected to have.

\section*{Acknowledgements}
We would like to thank Marc Verheijen and Adi Nusser for interesting  discussions  on  the  subject  presented  here.  We also  thank  the  Netherlands  Foundation  for  Scientific  Research support through the VICI grant 639.043.00.

\bibliographystyle{mnras}
\bibliography{HI.bib}

\appendix
\onecolumn
\section{Recombination and collisional ionization rates}\label{app:recion}
The ionization states of hydrogen and helium depend on the detailed balance between recombinations, ionizations and excitations. Below we list all the rates we adopted following \citet{art:fukkaw}.
\begin{itemize}
 \item Collisional ionization rates:\begin{itemize}
                               \item H I $\rightarrow$ H II:\begin{equation}
                                                        \beta_{\mathrm{HI}} = 5.85\times10^{-11}T^{1/2}\left(1+T_5^{1/2}\right)^{-1}\exp\left(-1.578/T_5\right)\quad \mathrm{cm^3\cdot s^{-1}}
                                                       \end{equation}
			       \item He I $\rightarrow$ He II:\begin{equation}
                                                        \beta_{\mathrm{HeI}} = 2.38\times10^{-11}T^{1/2}\left(1+T_5^{1/2}\right)^{-1}\exp\left(-2.853/T_5\right)\quad \mathrm{cm^3\cdot s^{-1}}
                                                       \end{equation}
			       \item He II $\rightarrow$ He III:\begin{equation}
                                                        \beta_{\mathrm{HeII}} = 5.68\times10^{-12}T^{1/2}\left(1+T_5^{1/2}\right)^{-1}\exp\left(-6.315/T_5\right)\quad \mathrm{cm^3\cdot s^{-1}}
                                                       \end{equation}
                              \end{itemize}
\item Recombination rates:\begin{itemize}
                               \item H II $\rightarrow$ H I:\begin{equation}
                                                        \alpha_{\mathrm{HII}} = 3.96\times10^{-13}T_4^{-0.7}\left(1+T_6^{0.7}\right)^{-1}\quad \mathrm{cm^3\cdot s^{-1}}
                                                       \end{equation}
			       \item He II $\rightarrow$ He I:\begin{equation}
                                                        \alpha_{\mathrm{HeII}} = 1.50\times10^{-10}T^{-0.6353}\quad \mathrm{cm^3\cdot s^{-1}}
                                                       \end{equation}
			       \item He III $\rightarrow$ He II:\begin{equation}
                                                        \alpha_{\mathrm{HeIII}} = 2.12\times10^{-12}T_4^{-0.7}\left(1+0.379T_6^{0.7}\right)^{-1}\quad \mathrm{cm^3\cdot s^{-1}}
                                                       \end{equation}
                              \end{itemize}
\item Dielectric recombination rate:\begin{itemize}
			       \item He II $\rightarrow$ He I:\begin{equation}
                                                        \xi_{\mathrm{HeII}} = 6.0\times10^{-10}T_5^{-1.5}\exp\left(-4.7/T_5\right)\left[1+0.3\exp\left(-0.94/T_5\right)\right]\quad \mathrm{cm^3\cdot s^{-1}}
                                                       \end{equation}
                              \end{itemize}
\end{itemize} 
\section{Cooling rates}\label{app:cool}
In this section we list the cooling rates that were included in our model.
\begin{itemize}
 \item Collisional ionization cooling:\begin{itemize}
                               \item H I:\begin{equation}
                                                        \zeta_{\mathrm{HI}} = 1.27\times10^{-21}T^{1/2}\left(1+T_5^{1/2}\right)^{-1}\exp\left(-1.58/T_5\right)\quad \mathrm{erg\cdot cm^3\cdot s^{-1}}
                                                       \end{equation}
			       \item He I:\begin{equation}
                                                        \zeta_{\mathrm{HeI}} = 9.38\times10^{-22}T^{1/2}\left(1+T_5^{1/2}\right)^{-1}\exp\left(-2.85/T_5\right)\quad \mathrm{erg\cdot cm^3\cdot s^{-1}}
                                                       \end{equation}
                               \item He II ($2^3$S):\begin{equation}
                                                        \zeta_{\mathrm{HeI,2^3S}} = 5.01\times10^{-27}T^{-0.1687}\left(1+T_5^{1/2}\right)^{-1}\exp\left(-5.53/T_4\right)n_{\mathrm{e}}n_{\mathrm{HeII}}/n_{\mathrm{HeI}}\quad \mathrm{cm^3\cdot s^{-1}}
                                                       \end{equation}                        
			       \item He II:\begin{equation}
                                                        \zeta_{\mathrm{HeII}} = 4.95\times10^{-22}T^{1/2}\left(1+T_5^{1/2}\right)^{-1}\exp\left(-6.31/T_5\right)\quad \mathrm{erg\cdot cm^3\cdot s^{-1}}
                                                       \end{equation}
                              \end{itemize}
\item Collisional excitation cooling:\begin{itemize}
                               \item H I:\begin{equation}
                                                        \psi_{\mathrm{HI}} = 7.5\times10^{-19}\left(1+T_5^{1/2}\right)^{-1}\exp\left(-1.18/T_5\right)\quad \mathrm{erg\cdot cm^3\cdot s^{-1}}
                                                       \end{equation}
			       \item He I:\begin{equation}
                                                        \psi_{\mathrm{HeI}} = 9.10\times10^{-27}T^{-0.1687} \left(1+T_5^{1/2}\right)^{-1} \exp\left(-1.31/T_4\right) n_{\mathrm{e}}n_{\mathrm{HeII}}/n_{\mathrm{HeI}}\quad \mathrm{erg\cdot cm^3\cdot s^{-1}}
                                                       \end{equation}
			       \item He II:\begin{equation}
                                                        \psi_{\mathrm{HeII}} = 5.54\times10^{-17} T^{-0.397}\left(1+T_5^{1/2}\right)^{-1} \exp\left(-4.73/T_5\right)\quad \mathrm{cm^3\cdot s^{-1}}
                                                       \end{equation}
                              \end{itemize}
                              
\item Recombination cooling:\begin{itemize}
                               \item H II:\begin{equation}
                                                        \eta_{\mathrm{HII}} = 2.82\times10^{-26}T_3^{0.3}\left(1+3.54T_6\right)^{-1}\quad \mathrm{erg\cdot cm^3\cdot s^{-1}}
                                                       \end{equation}
			       \item He II:\begin{equation}
                                                        \eta_{\mathrm{HeII}} = 1.55\times10^{-26}T^{0.3647}\quad \mathrm{erg\cdot cm^3\cdot s^{-1}}
                                                       \end{equation}
			       \item He III:\begin{equation}
                                                        \eta_{\mathrm{HeIII}} = 1.49\times10^{-25}T^{0.3}\left(1+0.855T_6\right)^{-1}\quad \mathrm{erg\cdot cm^3\cdot s^{-1}}
                                                       \end{equation}
                              \end{itemize}
\item Dielectric recombination cooling:\begin{itemize}
			       \item He II:\begin{equation}
                                                        \omega_{\mathrm{HeII}} = 1.24\times10^{-13}T_5^{-1.5} \left(1+0.3\exp\left(-9.4/T_4\right)\right)^{-1}\exp\left(-4.7/T_5\right)\quad \mathrm{erg\cdot cm^3\cdot s^{-1}}
                                                       \end{equation}
                              \end{itemize}
\item Free-free cooling:\begin{equation}
                         \theta_{\mathrm{ff}} = 1.42\times10^{-27}g_{\mathrm{ff}}T^{1/2}
                        \end{equation}
     With $g_{ff}$ = 1.1
\item Compton cooling:\begin{equation}
                       \lambda_{\mathrm{c}} = 4k_{\mathrm{B}}\left(T-T_{\gamma}\right)\frac{\pi^2}{15}\left(\frac{k_{\mathrm{B}}T_\gamma}{\hbar c}\right)^3\left(\frac{k_{\mathrm{B}}T_\gamma}{m_{\mathrm{e}}c^2}\right)n_{\mathrm{e}}\sigma_{\mathrm{T}}c
                      \end{equation}
      Where $T_\gamma$ is the temperature of the cosmic microwave background ($T_\gamma$ = 2.736(1+z) K).

\end{itemize} 

\section{Lyman alpha}\label{app:lya}
\begin{figure} 
\begin{centering}  
\includegraphics[angle=0,width=0.52\textwidth]{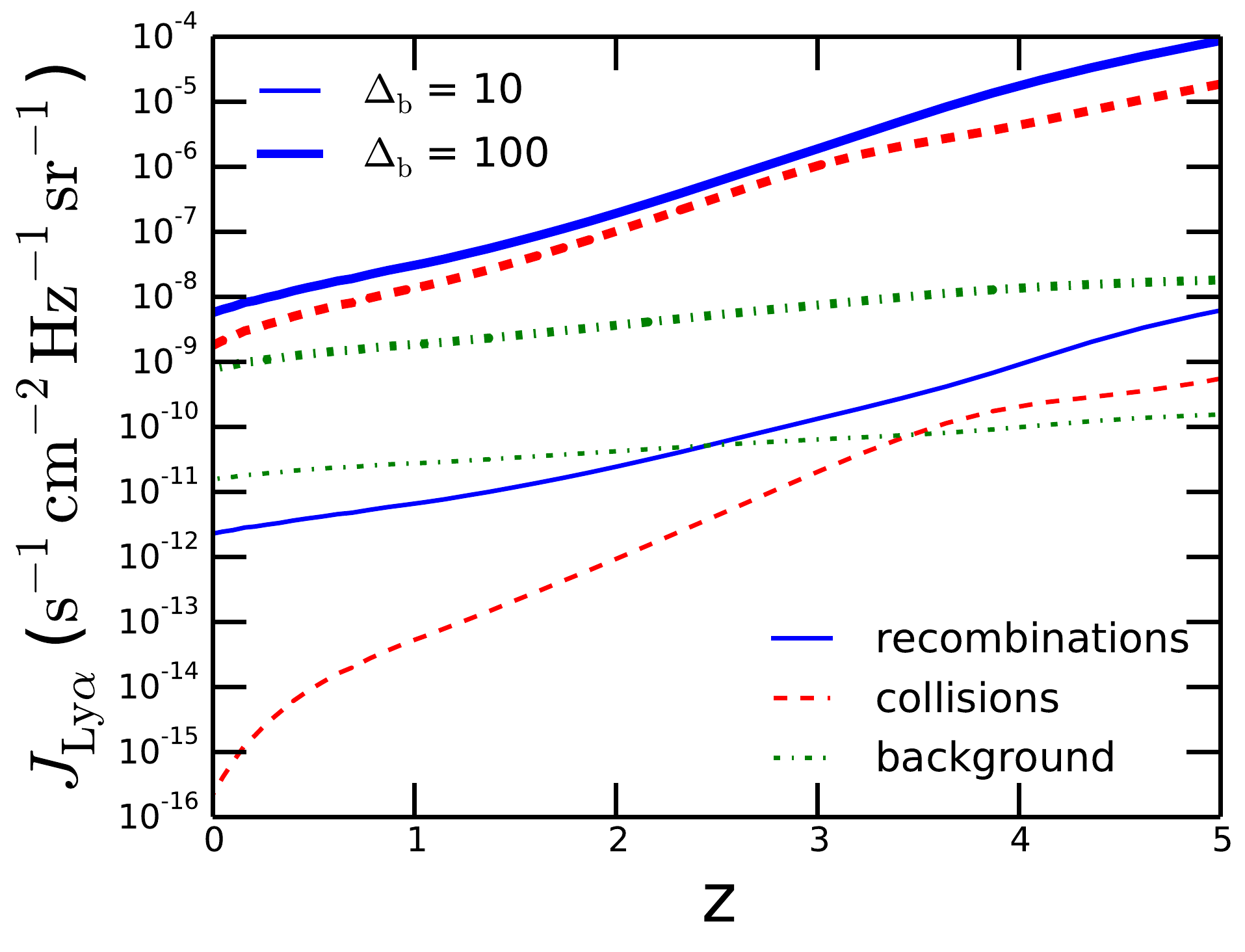}
\caption{Lyman alpha photon angle-averaged specific intensity for hydrogen for the different mechanisms. The linewidth denotes the density.}
\label{fig:jlya}
\end{centering}
\end{figure}

For the Wouthuysen-Field coupling it is necessary to assume a model of the Lyman alpha emission. We take into account three sources of Lyman alpha: collional excitations, recombinations and high energy photons from the X-ray/UV-background that redshift into the Lyman alpha line and then interact with the IGM. For the latter we adopt the \citet{art:hm12} model.\\
The Lyman alpha photon angle-averaged specific intensity (in units of $\mathrm{s}^{-1}\mathrm{cm}^{-2}\mathrm{Hz}^{-1}\mathrm{sr}^{-1}$) is calculated as follows
\begin{equation}
 J_{\mathrm{Ly\alpha,x}} = \frac{N_{\mathrm{Ly\alpha,x}}(z)D_\mathrm{A}^2}{4\pi D_\mathrm{L}^2}\frac{dr}{d\nu} = \frac{N_{\mathrm{Ly\alpha,x}}(z)}{4\pi}\frac{\lambda_{\mathrm{Ly\alpha,0}}}{H(z)}
\end{equation}
where the $D_{\mathrm{A}}$ and $D_{\mathrm{L}}$ are the angular and luminosity distances, respectively, and $\lambda_{\mathrm{Ly\alpha,0}} = 1215.76 \mbox{\AA}$ is the rest wavelength of the Lyman alpha transition. $N_{\mathrm{Ly\alpha,x}}$ is the number of Ly$\alpha$ photons that interact with the IGM per unit volume per unit time. For recombinations this number follows directly from the recombination rate
 \begin{equation}
  N_{\mathrm{Ly\alpha,rec}}(z) = f_{\mathrm{Ly\alpha}}\alpha_{\mathrm{HII}}\left(T_{\mathrm{k}},z\right)n_{\mathrm{e}}n_\mathrm{HII}\quad \mathrm{cm^{-3}s^{-1}}
 \end{equation}
where $f_{\mathrm{Ly\alpha}}$ is the fraction of recombinations that result in a Lyman alpha photon. This fraction depends on the temperature. We use the fitting function obtained by \citet{art:cantalupo}
\begin{equation}
  f_{\mathrm{Ly\alpha}}^{\mathrm{HI}} = 0.686 - 0.106\log T_4 - 0.009T_4^{-0.44},\label{eqrecfrac}
\end{equation}
which is accurate within 0.1$\%$ at temperatures 100 K $<$ T $<$ $10^5$ K. The recombination rate is given by
\begin{equation}
\alpha_{\mathrm{HII}} = 3.96\times10^{-13}T_4^{-0.7}\left(1+T_6^{0.7}\right)^{-1}(1+z)^3\quad \mathrm{cm^{3}s^{-1}}. \label{eqrecrate}
\end{equation}
Equations \ref{eqrecfrac} and \ref{eqrecrate} are the rates for case B recombination. The case A recombination rate is higher, but the corresponding probability of emitting a Lyman alpha photon per recombination is lower, effectively negating the difference \citep{art:dijkstra}.
Similarly, the density of Lyman alpha photons due to collisions that interact with the IGM follows from the collisional excitation coefficient $Q_{col}$ as
 \begin{equation}
  N_{\mathrm{Ly\alpha,col}}(z) = Q_{\mathrm{col}}\left(T_k,z\right)n_{\mathrm{e}} n_\mathrm{HI}\quad \mathrm{cm^{-3}s^{-1}}
 \end{equation}
 The comoving collisional excitation coefficient for transitions from the ground state to level $n$ is given by \citep{art:cantalupo}
\begin{equation}
 q_{1,n}^{\mathrm{HI}} = 8.629\times10^{-6}T^{-0.5}\frac{\Omega_n(T)}{\omega_{\mathrm{1}}}e^{E_n/k_{\mathrm{B}}T}(1+z)^3\quad \mathrm{cm^{3}s^{-1}},
\end{equation}
where $E_n$ is the energy corresponding to the transition, $\omega_{\rm{1}}$ is the statistical weight of the ground state and the function $\Omega_n$ is the effective collision strength, given by \citep{art:effcolstren}
\begin{equation}
 \Omega_n(T) = \begin{cases}
          3.44\times10^{-1} &+ 1.293\times10^{-5} T\\
          &+ 5.124\times10^{-12} T^2\\
          &+ 4.473\times10^{-17} T^3,\, n = 1\\
          5.462\times10^{-2} &- 1.099\times10^{-6} T\\
          &+ 2.457\times10^{-11} T^2\\
          &- 1.528\times10^{-16} T^3,\, n = 2\\
          4.838\times10^{-2} &+ 8.56\times10^{-7} T\\
          &- 2.544\times10^{-12} T^2\\
          &+ 5.093\times10^{-18} T^3,\, n = 3.
          \end{cases}
\end{equation}
The total collisional excitation coefficient is then the sum over all the collisional excitation coefficients, where we only consider transitions up to n = 3
\begin{equation}
 Q_{\mathrm{col}}^{\mathrm{HI}} = \sum^{3}_{n=1} q^{\mathrm{HI}}_{1,n}.
\end{equation}\\
Finally, for the background Lyman alpha emission there are two main contributions. At high redshift, quasars are dominant, whereas at low redshift the main contribution of Lyman alpha photons comes from the galaxies. For the quasars, the comoving emissivity at 1 ryd is \citep{art:hm12}

\begin{equation}
\epsilon_{912}(z) = \left(10^{24.6} \mathrm{erg\, s^{-1} Mpc^{-3} Hz^{-1}}\right)\times(1+z)^{4.68}\frac{\exp(-0.28z)}{26.3 + \exp(1.77z)},
\end{equation}
which is a fit to the \citet{art:hopkins} results. This is then integrated over frequency to get the quasar Lyman alpha photon density
 \begin{equation}
  N_{\mathrm{Ly\alpha,qso}} = \int_{\nu(z_{\mathrm{Ly\alpha}})}^{\nu(z_{\mathrm{max}})}\frac{\epsilon_{912}(z(\nu))}{h\nu}\left(\frac{\nu}{\nu_{912}}\right)^{-1.57}\mathrm{d}\nu,
 \end{equation}
where $z_{Ly\alpha}$ is the redshift at which the emission couples to the local IGM as Lyman alpha, $\nu_{912}$ is the frequency corresponding to 912 $\mbox{\AA}$, $\nu(z) = \nu_{Ly\alpha}(1+z)/(1+z_{\mathrm{Ly\alpha}})$ and the exponent -1.57 comes from the quasar UV SED for wavelengths below 1300 $\mbox{\AA}$.
The galactic contribution to the background Lyman alpha photons follows from a fit to the star formation rate density (SFRD) by \citet{art:hm12}

\begin{equation}
 SFRD(z) = \frac{6.9\times10^{-3}+0.14(z/2.2)^{1.5}}{1 + (z/2.7)^{4.1}}\mathrm{M_\odot yr^{-1}\mathrm{Mpc}^{-3}}
\end{equation}
To convert the observed luminosity densities $\rho_{1500\mbox{\AA}}$ to ongoing star formation rate densities, \citet{art:hm12} adopted a conversion factor $\kappa = 1.05\times10^{-28}$
\begin{equation}
 SFRD(t) = \kappa\times\rho_{1500\mbox{\AA}}(t),
\end{equation}
where $\rho_{1500\mbox{\AA}}$ is in units of erg $\mathrm{s}^{-1}\mathrm{Mpc}^{-3}\mathrm{Hz}^{-1}$. We adopt the same conversion factor to go from the fitted star formation rate density, back to a luminosity density. The galactic Lyman alpha photon density is then the integral over the SFRD
\begin{equation}
 N_{\mathrm{Ly\alpha,gal}} = \int_{\nu(z_{\rm Ly\alpha})}^{\nu(z_{\rm{max}})}\frac{SFRD(z(\nu))/\kappa}{h\nu}\frac{B_\nu(T_\mathrm{gal})}{B_{\mathrm{1500}}(T_\mathrm{gal})}\mathrm{d}\nu
\end{equation}
where $B_\nu$ is the Planck function at frequency $\nu$ and $B_\mathrm{1500}$ is the same at 1500 $\mbox{\AA}$. The Planck functions depend on the temperature of the IGM in the galaxies, for which we adopt $T_{\mathrm{gal}}$ = 6000 K.\\
One more effect that needs to be taken into account is that photons can scatter multiple times before they redshift out of the Lyman alpha line. Each additional scattering adds to the Wouthuysen-Field coupling and so the Wouthuysen-Field coupling factor $x_{\rm c}$ needs to be multiplied by the number of scatterings\citep{art:field59}. This is given by the Gunn-Peterson optical depth \citep{art:GP} 
\begin{equation}
 \tau_{\mathrm{GP}} = \frac{n_\mathrm{x}\lambda_\mathrm{Ly\alpha}}{H(z)}\frac{f_\alpha\pi e^2}{m_\mathrm{e}c}
\end{equation}
with $n_{\rm{x}}$ either the HI or $\rm{^3HeII}$ density. Because Lyman alpha photons created through recombinations originate close to the line center, their Lyman alpha specific intensity gets an additional factor of 1.5 \citep{art:field59}.\\
The evolution of the Lyman alpha intensity is shown in Figure \ref{fig:jlya}. At low redshift, for the higher densities recombination and collisions are the dominant source of Lyman alpha, whereas for the lower densities it is the background emission.

\end{document}